\documentclass[reprint,superscriptaddress,aip,jcp,longbibliography]{revtex4-2}
\usepackage{graphicx}
\usepackage{amsmath,amssymb}
\usepackage{booktabs}
\usepackage{placeins}
\newcommand{\papertitle}{A general-purpose electron density model for molecular chemistry}
\newcommand{\PETrho}{\mbox{PET-rho-OMol25}}
\newcommand{\PETr}{\mbox{PET-rho}}

\begin{document}
\title{\papertitle}
\author{Paolo Pegolo}
\affiliation{Laboratory of Computational Science and Modeling, Institut des Mat\'eriaux, \'Ecole Polytechnique F\'ed\'erale de Lausanne, 1015 Lausanne, Switzerland}

\author{Joseph W. Abbott}
\affiliation{Laboratory of Computational Science and Modeling, Institut des Mat\'eriaux, \'Ecole Polytechnique F\'ed\'erale de Lausanne, 1015 Lausanne, Switzerland}

\author{Michele Ceriotti}
\email{michele.ceriotti@epfl.ch}
\affiliation{Laboratory of Computational Science and Modeling, Institut des Mat\'eriaux, \'Ecole Polytechnique F\'ed\'erale de Lausanne, 1015 Lausanne, Switzerland} \date{\today}
\begin{abstract}
Universal machine-learning potentials predict molecular structure and stability at the accuracy of quantum mechanics, for a fraction of the cost. Properties derived instead from knowledge of the electronic structure give profound insights into chemical reactivity, but are not accessible with the same ease or transferability. 
We train PET-rho-OMol25, a model of the ground-state electron density and spin polarization of molecular systems on four million OMol25 structures in multiple charge and spin states.
The model makes predictions for entire proteins in a few seconds on a single GPU, and enables estimation of important diagnostics of molecular structure and reactivity, such as vibrational Stark measurements, halogen $\sigma$-holes, hydrogen-bond basicities and core-level shifts, as well as spin density for redox chemistry and magnetism.
PET-rho-OMol25 has an accuracy close to density-functional theory, with its least reliable predictions flagged by its built-in uncertainty estimator, for most properties and classes of compounds.
It can be used as a drop-in replacement for electronic-structure calculations of the density, making insightful analyses available for larger and more complex structures. \end{abstract}
\maketitle

The ground-state electron density determines all properties of an electronic system~\cite{hohenberg1964inhomogeneous}, including the electrostatic potential (ESP) that influences how a molecule is solvated, binds to other molecules and takes part in catalysis~\cite{warshel2006electrostatic,fried2017review}.
These processes often involve thousands of atoms, as in an enzyme or a solvated complex, sizes for which even density-functional theory (DFT), one of the most affordable first-principles routes to the density, becomes impractical.
Machine-learning (ML) models trained on DFT data of small structures instead predict the density with an accuracy close to that of their reference, at a cost that grows linearly with the number of atoms~\cite{grisafi2018transfer,fabrizio2019noncovalent,lewis2021condensed,briling2021metrics,rackers2023cracking,elsborg2025electra}.
ML models of atomic multipoles give the ESP at a similar cost~\cite{bereau2015multipole,thurlemann2022multipole}, but not its short-range part, where the charge clouds of neighboring molecules overlap.

These models are often trained on a single class of systems, such as bulk materials and interfaces~\cite{jorgensen2022deepdft,focassio2023jacobi,koker2024charge3net,grisafi2023metal,grisafi2024qmmm,ellis2021mala,fiedler2023mala,elsborg2025electra,qin2026eacnet,li2025edengnn,zeng2026aiden,elsborg2026electrafi} or gas-phase organic molecules and their complexes~\cite{brockherde2017bypass,fabrizio2019noncovalent,cuevaszuviria2021a2md,rackers2023cracking}, so that a new model must be trained on new DFT data for each new class of chemistry (SI, Table~\ref{tab:prior}).
The OMol25 dataset~\cite{omol} provides around four million DFT densities (its `4M' split) that cover most of molecular chemistry, including charged systems and molecules with different spin multiplicities.
Training on a dataset of this size requires a model that is both expressive and fast.
The point-edge transformer (PET)~\cite{pozdnyakov2023pet} is an unconstrained graph neural network that learns rotational equivariance from data augmented with rotations~\cite{domina2026unconstrained} instead of imposing it, which makes it flexible and cheap to evaluate.
We recently showed that it gives a convenient balance of speed, accuracy, and memory footprint for the density and other tensorial properties~\cite{abbott2026unconstrained}.
Here, we train PET on OMol25 to obtain a general-purpose model of the electron density for molecules in the charge and spin states covered in the dataset, with an estimate of the error of each prediction that needs no reference calculation.
We compare its error with the lowest error that its basis allows, test it on four electrostatic observables that can be compared with experiment, and identify where it fails and whether its error estimate flags those failures.

\begin{figure}[!t]
    \centering
    \includegraphics{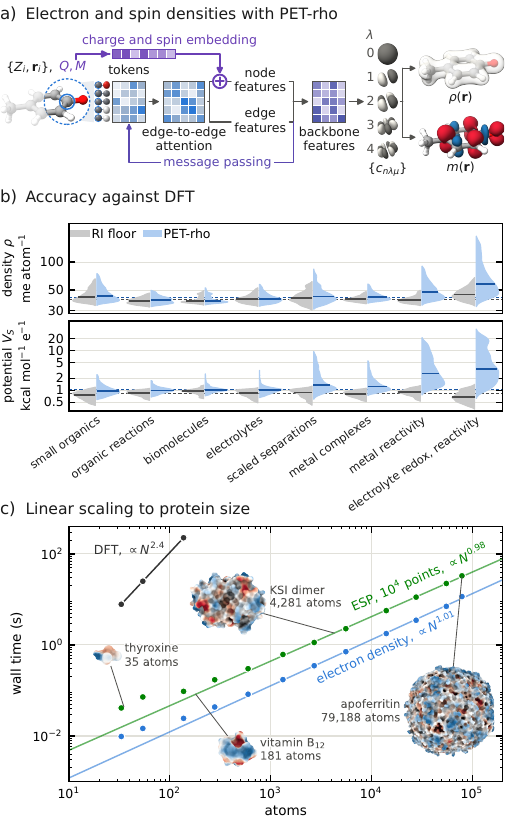}
    \caption{%
    Architecture, accuracy, and cost of \PETr{}.
    (a)~The \PETr{} backbone produces node and edge features, with the charge $Q$ and multiplicity $M$ embedding summed with the former. The backbone features map to the coefficients of the electron density $\rho$ and spin density $m$.
    (b)~Error against DFT for eight OMol25 subsets (Table~\ref{tab:nmae-category}): integrated density error per atom and root-mean-square ESP error on the 0.001~a.u.\ isodensity surface (39,845 structures).
    Violins show the central 90\% of structures for \PETr{} (blue) and for the fitted density, the RI floor (grey), with medians as dark lines.
    Dashed lines mark the medians over all test structures.
    (c)~Wall time on one NVIDIA GH200 (mean of three repeats), with the ESP on the molecular surface in the insets (red negative, blue positive). DFT is $\omega$B97M-V/def2-TZVPD in GPU4PySCF~\cite{li2025gpu4pyscf} with density fitting of Coulomb and exchange, from setup to convergence of the self-consistent field (SCF); small systems are batched for \PETr{}.}
    \label{fig:accuracy}
\end{figure}

OMol25 contains over 100 million $\omega$B97M-V/def2-TZVPD~\cite{mardirossian2016wb97mv,weigend2005balanced,rappoport2010property} calculations of systems with up to 350 atoms and covering 83 elements, from small organic molecules to biomolecular fragments, electrolytes and metal complexes~\cite{omol}.
Its 4M split (3,986,754 structures) includes the full DFT output, from which we take the electron and spin densities.
To keep the model's target atom-centered, it is convenient to expand the density on a localized atomic basis, as in the resolution-of-the-identity (RI) approximation~\cite{vahtras1993integral}. This is well-suited for an atomistic model based on local features, since these models can be trained on small systems and applied to larger ones, from large complexes to the condensed phase~\cite{grisafi2018transfer,lewis2021condensed,grisafi2022electronic,rackers2023cracking,liu2026scfbench,abbott2026unconstrained}.
As atomic orbitals, we choose the \texttt{def2-universal-jfit} auxiliary basis, optimized to fit the Coulomb energy of any def2 orbital basis for the elements from H to Rn~\cite{weigend2006coulomb}.
With a few tens of functions per atom (e.g., 49 for carbon, 89 for iron), this basis compactly describes the density for all the elements of OMol25.
The fit uses the Coulomb metric at fixed electron count, which minimizes the electrostatic self-energy of the fitting error~\cite{whitten1973coulombic}.
Any auxiliary basis is incomplete, so even the best fit does not exactly represent the full DFT density.
A model trained on the fitted coefficients can at best reproduce the fitted density, so the error of the fit against DFT (the `RI floor') is, in the Coulomb metric, the lowest error the model can reach.

We introduce \PETrho{} (from here on referred to as \PETr{}), a general-purpose model for the electron (and spin) density of molecular systems (Fig.~\ref{fig:accuracy}a).
Because OMol25 contains charged and open-shell systems, the total charge and the spin multiplicity enter \PETr{} as global embeddings~\cite{bigi2026pushing,omol}, so the same geometry can be evaluated in any charge and spin state covered in the training set (Computational Methods).
The predicted coefficients are then shifted by the smallest change, in the Coulomb metric, that gives the density the exact number of electrons and the spin density the exact number of unpaired electrons~\cite{briling2021metrics}, so that the net charge of every predicted density, and hence its long-range ESP, is exact.

In Fig.~\ref{fig:accuracy}b, we compare the error of both \PETr{} and the RI floor against DFT, for eight OMol25 subsets (SI, Table~\ref{tab:nmae-category}).
Since the floor is the lowest error the model can reach, the distance between the two distributions shows the share of the error that comes from the model.
Over all 39,845 test structures, the median of the integrated absolute density error of \PETr{} is 41 millielectrons (me) per atom (1\% of the electrons) against 40\,me per atom for the fitted density, indicating that for a typical structure the error comes almost entirely from the basis.
This is the case for small organic molecules, organic reactions, biomolecules and electrolytes, where \PETr{} is as good as the RI floor both for the density and for the ESP on the molecular isodensity surface ($\rho = 0.001$~a.u.), which is within about 1\,kcal\,mol$^{-1}$\,e$^{-1}$ of DFT.
In the `metal reactivity' and `electrolyte redox and reactivity' subsets, instead, \PETr{} stays above the floor, by 20--40\% for the density and by a factor of three to eight for the ESP (Fig.~\ref{fig:accuracy}b and SI, Sec.~\ref{sec:si-esp}), indicating that in these two subsets the accuracy is limited by the model itself.
In every subset, including these two, the ESP of \PETr{} is closer to DFT than that of the density of the semiempirical GFN2-xTB~\cite{bannwarth2019gfn2}, by a factor of about three in metal reactivity and of four to five in the others (SI, Sec.~\ref{sec:si-baselines}).
Because the density is predicted from local features, the cost of a prediction grows linearly with the number of atoms, so that a whole protein assembly, the apoferritin 24-mer with nearly 80,000 atoms, takes 12 seconds on a single NVIDIA GH200 GPU.
In the same wall time, DFT at this level of theory converges only for molecules more than three orders of magnitude smaller (Fig.~\ref{fig:accuracy}c; SI, Sec.~\ref{sec:si-models}).

\begin{figure}[!t]
    \centering
    \includegraphics{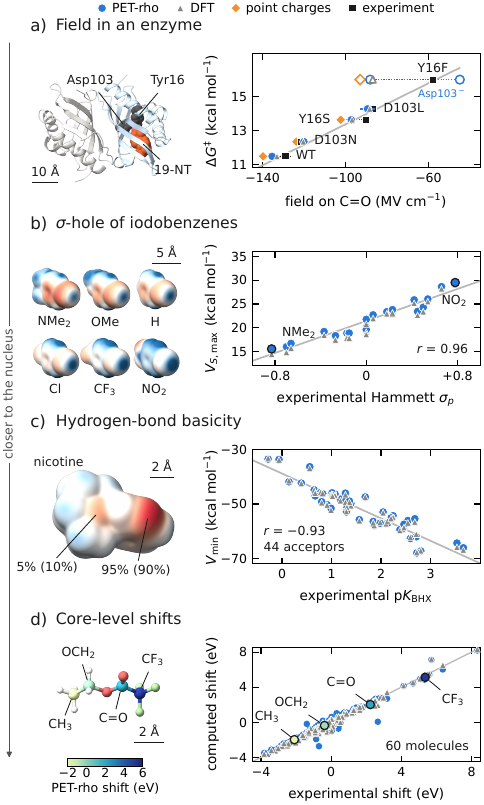}
    \caption{%
    Observables from the \PETr{} density, with DFT and experiment.
    (a) Activation free energy of KSI variants against the field on the substrate C=O bond~\cite{fried2014ksi}, with DFT as the \PETr{} field plus the mean DFT--\PETr{} difference on active-site clusters of each variant; open symbols mark Y16F, for which the sampled structure is likely far from the experimental one (see text), and the open circle labelled Asp103$^-$ is Y16F with Asp103 ionized, while D40N, in which the general base is removed, is omitted as in the original study~\cite{fried2014ksi}.
    (b) $\sigma$-hole maximum of \textit{para}-X-iodobenzenes against the Hammett constant~\cite{hansch1991survey}. $r$ denotes the Pearson correlation coefficient.
    (c) Lone-pair ESP minimum of hydrogen-bond acceptors against p$K_\mathrm{BHX}$~\cite{laurence2009pkbhx}, and for nicotine the predicted (measured~\cite{graton2003nicotine}) share of complexes at each nitrogen.
    (d) Core-level shifts against experiment for 142 sites in 60 molecules, with one offset per element over all sites, and the carbons of ethyl trifluoroacetate colored by their shift.
    Renders show the ESP on the 0.001~a.u.\ isodensity surface (red negative; $\pm 20$ and $\pm 40$~kcal\,mol$^{-1}$ in b and c); details in SI, Secs.~\ref{sec:si-ksi}--\ref{sec:si-xps}.}
    \label{fig:observables}
\end{figure}

To understand how useful such a model can be in practice, it is necessary to look at properties that depend on the density as a scalar field, going beyond what the integrated density error can indicate.
We therefore test \PETr{} on four electrostatic properties that can be compared with experiment: the field an enzyme exerts on its substrate, the $\sigma$-hole of halogen-bond donors, the hydrogen-bond basicity of acceptors, and the core-level shifts measured by X-ray photoelectron spectroscopy (XPS) (Fig.~\ref{fig:observables}).
The first is computed on a system far too large for DFT, while the other three are computed on small molecules, for which DFT is also affordable.
As cheaper alternatives to DFT that can serve as baseline, we compare with the point charges of a protein force field for the enzyme and with the density and atomic charges of GFN2-xTB for the other properties.

The catalytic effect of enzymes is partly determined by the electric field that they exert on the substrate~\cite{warshel2006electrostatic,fried2017review}.
In ketosteroid isomerase (KSI), this field can be measured on the C=O bond of the substrate through its vibrational Stark shift~\cite{fried2014ksi}.
Across the wild-type enzyme and a series of mutants, in which single residues of the active site are replaced, the measured field correlates with the activation free energy of the reaction~\cite{fried2014ksi}.
Computing this field from DFT is out of reach, since it depends on the charge density of thousands of atoms and must be thermodynamically sampled to be compared with experiment.
We therefore sample the KSI dimer bound to the steroid 19-nortestosterone, for the wild type and five mutants, by machine-learning/molecular-mechanics (ML/MM) dynamics (Computational Methods), where residues within 10\,\AA{} of the ligand are described by the PET-OMol-L interatomic potential~\cite{bigi2026pushing} (also trained on OMol25), and the rest of the protein and the solvent by a force field. Simulating the whole system with an OMol25-trained potential would also be feasible~\cite{gao2026enzyme}, but the ML/MM approach is more efficient at sampling the ligand region of interest (SI, Sec.~\ref{sec:si-ksi}).
Along these trajectories, the field on the C=O bond is computed from the \PETr{} density of the whole periodic cell, about 40,000 atoms of protein, water and ions, which takes about 10~s per frame on a GH200 GPU (SI, Sec.~\ref{sec:si-ksi}).

In Fig.~\ref{fig:observables}a, we plot the measured activation free energy of each KSI variant~\cite{fried2014ksi} against the electric field, computed with the models or measured experimentally, and the line is a linear fit of the measured activation free energies against the measured fields.
Over the five variants of Fig.~\ref{fig:observables}a, the field of \PETr{} differs from the measured one by 10~MV\,cm$^{-1}$ on average (4~MV\,cm$^{-1}$ without the Y16F variant, see the discussion below), and that of the point charges of the force field~\cite{cornell1995ff94,jorgensen1983tip3p} on the same trajectories by 13~MV\,cm$^{-1}$ (7~MV\,cm$^{-1}$).
Since both are evaluated on the same structures, this difference comes from the electrostatics.
To check the density itself, we cut 36 clusters of the active site from earlier ML/MM runs with a smaller ML region, small enough for DFT (SI, Sec.~\ref{sec:si-ksi}).
On these clusters, the field of \PETr{} is within 1.6~MV\,cm$^{-1}$ of DFT on average, against 3.8~MV\,cm$^{-1}$ for the point charges, and \PETr{} also reproduces how Tyr16 and Asp103 are polarized by the rest of the active site.
Correcting the field of each variant by the mean difference between DFT and \PETr{} on its clusters moves it by at most 2.4~MV\,cm$^{-1}$ (triangles in Fig.~\ref{fig:observables}a).
Computations for the Y16F variant (open symbols) do not agree with the experiments, with a field about 30~MV\,cm$^{-1}$ too strong for both \PETr{} and point charges.
This discrepancy was also observed in the original experimental study, where point charges on the crystal structure accounted for only a third of the field lost with the Tyr16 hydroxyl~\cite{fried2014ksi}.
For Y16F, as for the other variants, the cluster fields agree with DFT and the computed C=O frequencies follow the computed field, so the deviation likely comes from the sampled structure (SI, Sec.~\ref{sec:si-ksi}).
One possibility that would move the field closer to experiment is a different protonation state of Asp103, which has not been measured in Y16F. With Asp103 ionized (Asp103$^-$ in Fig.~\ref{fig:observables}a), the field of Y16F is 13~MV\,cm$^{-1}$ weaker than measured, against 30~MV\,cm$^{-1}$ stronger with Asp103 neutral.
Computing the p$K_\mathrm{a}$ of Asp103 in the enzyme, for example from the free-energy profile of the proton dissociation~\cite{xu2026acidbase,kabylda2026so3lr}, would test this hypothesis.
In general, computed properties of enzymes with OMol25-trained potentials can be similarly sensitive to the sampled structure, as shown for reaction barriers in chorismate mutase~\cite{gao2026enzyme}, but this sensitivity also suggests that screening structural changes for experimental agreement in the electrostatics becomes accessible with an accurate surrogate model of the density.

Besides the field inside a protein, the ESP on the surface of a molecule indicates where and how strongly other molecules bind to it.
A paradigmatic case is the halogen bond, which is often designed into drug candidates to bind them to their protein targets~\cite{wilcken2013halogen}.
On a halogen bonded to carbon, the ESP is positive along the C--X bond and negative around it, so that the halogen binds an electron donor on the side opposite to the carbon~\cite{clark2007sigmahole,politzer2013sigmahole}.
This positive cap, the `$\sigma$-hole', cannot be described by a charge on the halogen, which can only give the same ESP in every direction.
The strength of the $\sigma$-hole is commonly tuned with substituents on an aromatic ring, as in the 20 \textit{para}-X-iodobenzenes of Fig.~\ref{fig:observables}b.
With \PETr{}, the $\sigma$-hole grows linearly with the Hammett constant of X, an experimental measure of how strongly X withdraws electrons~\cite{hammett1937effect,hansch1991survey}, from 16 to 30~kcal\,mol$^{-1}$ between NMe$_2$ and NO$_2$, and is 0.4 to 1.2~kcal\,mol$^{-1}$ above DFT (SI, Sec.~\ref{sec:si-sigma}).
The density and the atomic charges of GFN2-xTB follow the same trend, but their ESP on the cap of iodine is negative, so they predict no $\sigma$-hole and hence no halogen bond (SI, Sec.~\ref{sec:si-baselines}).

From the ESP, we can also estimate which atom of a molecule accepts a hydrogen bond, and how strongly, through the minimum near each lone pair, $V_\mathrm{min}$~\cite{kenny1994hbond,kenny2016hbond}.
A comprehensive measure of the acceptor strength is given by the p$K_\mathrm{BHX}$ hydrogen-bond basicity scale~\cite{laurence2009pkbhx}.
Over 44 acceptors in seven families, from ethers to phosphine oxides, which we selected from a public compilation of measurements~\cite{kenny2020pkbhxdata}, $V_\mathrm{min}$ of \PETr{} is within 0.5~kcal\,mol$^{-1}$ of DFT on average.
It correlates with p$K_\mathrm{BHX}$ as well as the $V_\mathrm{min}$ of DFT ($r = -0.93$ and $-0.94$) and better than that of the GFN2-xTB density ($r = -0.77$; Fig.~\ref{fig:observables}c; SI, Secs.~\ref{sec:si-hbond} and \ref{sec:si-baselines}).
For molecules with more than one acceptor, the fit of $V_\mathrm{min}$ to p$K_\mathrm{BHX}$ can be used to estimate the share of hydrogen-bonded complexes at each site, as for nicotine, where a proton binds to the pyrrolidine nitrogen (the more basic of the two) yet 90\% of the hydrogen bonds with 4-fluorophenol form on the pyridine nitrogen~\cite{graton2003nicotine}.

Inside a molecule, the ESP at the nuclei is related to the binding energy of core electrons, which can be probed experimentally via XPS.
The shift of this binding energy between atoms of the same element is used to identify their chemical environment, often by comparison with shifts computed for candidate structures, which can be done with machine-learning models~\cite{elmachachi2024graphene,hernandez2026lixsi}.
It amounts to a few eV, for instance 7.5\,eV between the CH$_3$ and CF$_3$ carbons of ethyl trifluoroacetate (Fig.~\ref{fig:observables}d).
The CF$_3$ carbon has fewer electrons around it than the CH$_3$ carbon, so the ESP at its nucleus is more positive and its core electrons are more strongly bound.
Once a core electron is removed, the valence electrons relax towards the hole and lower the binding energy.
In the equivalent-cores approximation, the hole acts on the valence electrons as one extra charge on the nucleus, so that the ionized atom can be replaced by a cation of the next element of the periodic table~\cite{jolly1970equivalent}.
The shift of the binding energy is then given by the mean of the ESP at the nucleus in the neutral molecule and in this cation~\cite{davis1972relaxation}, with both densities predicted by \PETr{}.
The auxiliary basis describes the density close to the nuclei only roughly, but the core density is nearly the same at all sites of an element.
Replacing a DFT density by its fit on this basis changes the C1s shifts by at most 0.07~eV (SI, Sec.~\ref{sec:si-xps}).
Since only shifts are computed in this approximation, we align the binding energies of each element with experiment through one constant, fitted over all its sites, as in the potential models of core-level shifts~\cite{gelius1974binding}.
Over 142 sites in 60 molecules, \PETr{} is then within 0.24~eV of gas-phase experiment on average.
In the same approximation, DFT is within 0.17~eV, while with the density of GFN2-xTB the error is 1.9~eV (SI, Sec.~\ref{sec:si-baselines}).
The largest errors of \PETr{}, up to 2.6~eV, come from equivalent-core cations far from the structures of OMol25, such as NO$^+$ for the carbon of CO and CH$_3$CO$^+$ for the nitrogen of acetonitrile.
At the nuclei of the neutral molecules, the ESP of \PETr{} is within 0.10~eV of DFT at every site (SI, Sec.~\ref{sec:si-xps}).
As we show below, these cations are also flagged by the uncertainty estimate of the model.

\begin{figure}[t]
    \centering
    \includegraphics{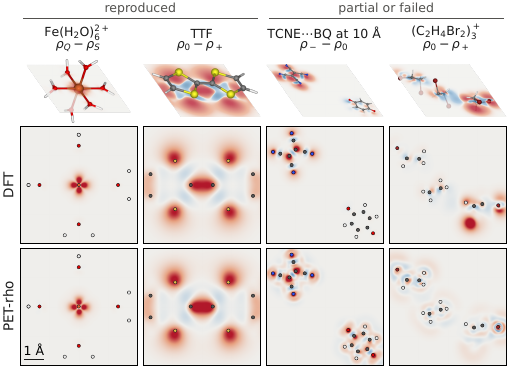}
    \caption{%
    Change of the total density between two states at one geometry, from DFT (top) and \PETr{} (bottom) in one plane per system, shown in the render above each column: Fe(H$_2$O)$_6^{2+}$ from singlet to quintet, the tetrathiafulvalene (TTF) cation, the anion of TCNE and benzoquinone (BQ) 10~\AA{} apart, and the cation of three 1,2-dibromoethane molecules; subscripts S and Q denote the singlet and quintet, and 0, $+$ and $-$ the neutral, cationic and anionic states.
    Color scales span $\pm$ the 99.5th percentile of the absolute DFT map (red positive, blue negative): from left to right, 0.61, 0.0085, 0.011 and 0.0074~e\,bohr$^{-3}$.
    Dots mark atoms within 0.9~\AA{} of the plane.}
    \label{fig:spin}
\end{figure}

The systems for which we tested these four observables are closed-shell and belong to the subsets in which \PETr{} reaches the RI floor.
The two other subsets, metal reactivity and electrolyte redox and reactivity, consist largely of open-shell structures along reaction paths or in oxidized and reduced states, which are also the hardest for interatomic potentials trained on OMol25~\cite{omol}.
These structures are difficult for DFT as well, and in fact for some of them the reference calculations converge only slowly and, depending on the initial guess, can end in different solutions (SI, Sec.~\ref{sec:si-test}). This means that the reference density is less well defined than for a closed-shell molecule.
For these structures, \PETr{} is about as far from the reference as the solutions are from each other, so part of its lost accuracy in these subsets likely comes from the reference rather than from the model.

Integrated errors such as those of Fig.~\ref{fig:accuracy} do not show whether spin density localizes at the correct molecular site, which is often the determining factor for redox chemistry or magnetism.
We therefore assess the change of the total density between two charge or spin states at one geometry, which shows where the added, removed or rearranged electrons go (Fig.~\ref{fig:spin}).
Within a molecule, the charge and spin distribution from \PETr{} is consistent with DFT (except for the negative spin populations produced by spin polarization, which are recovered at about a third of their magnitude; see SI, Sec.~\ref{sec:si-charge-spin}).
When the model has to choose where to place charge between molecules, however, it often fails.
With tetracyanoethylene (TCNE) and benzoquinone 10~\AA{} apart, the excess electron of the anion is placed entirely on TCNE---the stronger acceptor---within DFT, whereas with \PETr{} only 0.62\,e end up on TCNE, while the rest is spread over benzoquinone.
Similarly, in the cation of three 1,2-dibromoethane molecules, the hole should sit on one molecule, whereas with \PETr{} it is spread over all three, even though the training set contains other geometries of this cation (SI, Sec.~\ref{sec:si-charge-spin}).
The placement likely fails because the charge and the multiplicity enter as a single global embedding, added to the node features of every atom, so that separate molecules receive the same information about the presence of an excess electron, irrespective of their electron affinities.

\begin{figure}[t]
    \centering
    \includegraphics{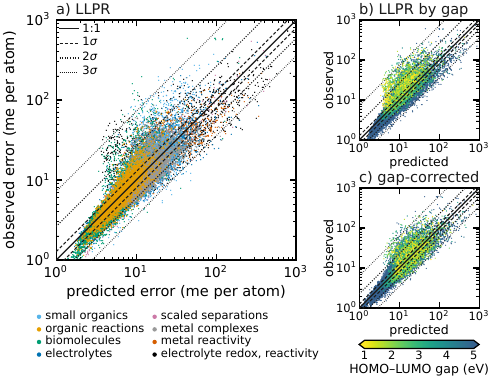}
    \caption{%
    Integrated absolute error of the \PETr{} electron density per atom, against the density fitted to DFT, for the 39,845 test structures, compared with the error predicted by the last-layer prediction rigidity (LLPR) ensemble (Computational Methods).
    (a)~Colored by chemistry subset as in Fig.~\ref{fig:accuracy}.
    (b)~The same structures colored by HOMO--LUMO gap.
    (c)~As (b), with the LLPR estimate multiplied by the median ratio of observed to predicted error of the validation structures of similar predicted gap.
    Axes of (b) and (c) are the same as (a).
    Solid lines: observed equal to predicted; dashed, dotted and fine dotted lines: bands around it with the widths of the 1$\sigma$, 2$\sigma$ and 3$\sigma$ quantiles of observed/predicted on the validation set, before (a, b) and after (c) the gap correction.
    Other scalings are compared in Fig.~\ref{fig:si-trust-alpha}.}
    \label{fig:trust}
\end{figure}

Given that, at inference, the DFT reference is not available, it is useful to know how reliable the predictions of the model are.
The last-layer prediction rigidity (LLPR)~\cite{bigi2024llpr,chong2025} formalism provides an estimate of the model uncertainty from a second-order (Laplace) approximation of the loss as a function of the last linear layer~\cite{mackay1992practical}, which can also be cast as the spread of a shallow, last-layer ensemble. This spread is wider for structures that the model sees as unlike its training set~\cite{bigi2024llpr}.
We sample 32 last layers from this approximate posterior, with the covariance expressed in the Coulomb metric of the auxiliary basis, and rescale the deviations of each block of element $Z$ and angular momentum $\lambda$ by a factor fitted on the validation set (Computational Methods).
The predicted error of a structure is the integrated absolute deviation of the member densities from the \PETr{} density, averaged over the ensemble, and is directly comparable with the observed error against the fitted density.
The bands in Fig.~\ref{fig:trust} are the empirical 68, 95 and 99.7\% intervals (1, 2, and 3$\sigma$) of the ratio between observed and predicted errors on the validation set.
Most test structures lie within the 2$\sigma$ band (Fig.~\ref{fig:trust}a), but a minority lies above it, mostly in the `electrolyte redox and reactivity' and `metal reactivity' subsets, where the error is underestimated by about a factor of two. Coloring the points by their HOMO--LUMO gap shows that these are mostly structures with a small gap (Fig.~\ref{fig:trust}b). Their density likely depends strongly on the occupation of the frontier orbitals, yet their last-layer features need not be uncommon with respect to the training set, so their LLPR uncertainty remains small.

The LLPR estimate can be used as a flag, e.g., by placing the predicted error of a new structure among those of the test set and flagging it above the 90th percentile.
The KSI clusters, the iodobenzenes and the hydrogen-bond acceptors of Fig.~\ref{fig:observables} all fall at or below the 52nd percentile of the test set, in line with their small differences from DFT.
Flagged structures appear only in the core-level benchmark, among them all the equivalent-core cations with neon (rare in OMol25), and in fact the 33 sites whose molecule or cation is flagged have a mean error against experiment of 0.49\,eV, against 0.16\,eV for the other 109 (SI, Sec.~\ref{sec:si-uq}).
To correct the overconfidence at small gap, we train an additional head on the frozen \PETr{} backbone to predict the gap via extremal pooling~\cite{malosso2026transferable}, and rescale the LLPR estimate by the median ratio of observed to predicted error of validation structures with a similar predicted gap~\cite{guan2023localized}.
For the test structures with gaps below 2\,eV, this ratio drops from 1.8 to 1.1 (Fig.~\ref{fig:trust}c; SI, Sec.~\ref{sec:si-uq}).
No such correction is possible for the charge placed between molecules in Fig.~\ref{fig:spin}: since the charge enters \PETr{} as a global embedding, the features of each fragment carry no information on where it goes, and the predicted error of these systems stays in the bulk of the test set.

In summary, we have presented \PETrho{}, a model of the electron (and spin) density of molecules trained on about four million structures of OMol25 in all their charge and spin states.
On most of the covered chemistry, the model-predicted coefficients match DFT densities as well as the reference ones, at a linear cost in the number of atoms. Computing the density of an 80,000-atom system takes only a few seconds on one GH200 GPU, making ab initio-quality electrostatic properties accessible on systems far beyond the reach of DFT.
We compared electrostatic properties computed from this density with four experimental observables: the field on the substrate of KSI along ML/MM trajectories, the $\sigma$-holes of 20 iodobenzenes against their Hammett constants, the hydrogen-bond basicity p$K_\mathrm{BHX}$ of 44 acceptors, and core-level binding energies measured by XPS.
For KSI, the field is marginally closer to experiment than that of the point charges of a protein force field on the same trajectories.
For the other three, where quantum chemistry is affordable, the agreement is close to that obtained with the DFT density, and better than with the density of GFN2-xTB.

The model comes with a built-in uncertainty estimate from an LLPR ensemble, which flags the structures with the largest errors and, once rescaled with a predicted HOMO--LUMO gap, also covers the small-gap structures of metal and electrolyte chemistry.
Placement of an excess electron or hole that can localize on more than one site, however, remains a limitation, because the charge enters the model as a single global embedding, and the uncertainty estimate does not flag it.
The same limitation may affect interatomic potentials that are conditioned on the total charge in the same way, whose treatment of charge transfer remains largely untested~\cite{gao2026enzyme}.
Placing the charge correctly without system-specific fitting requires a model in which the charge responds to the environment, for instance through charge equilibration~\cite{ko2021fourth}, which we leave to future work.

\section*{Computational Methods}
\paragraph{Data.}
The total and spin density matrices of OMol25 were fitted on the \texttt{def2-universal-jfit} auxiliary basis in the Coulomb metric~\cite{whitten1973coulombic}, with the number of electrons (of unpaired electrons) fixed.
The fitted coefficients are the training targets.
The 4M split of OMol25 was divided into training, validation and test sets in a 98:1:1 ratio, stratified by subset. The split is random by structure, so other geometries of a test structure can appear in the training set.
The density error of a structure, $\int |\rho_\mathrm{pred} - \rho_\mathrm{ref}|\,d\mathbf{r}$, was evaluated on a PySCF integration grid against the fitted density and against the DFT density, and is reported per atom (in SI, Sec.~\ref{sec:si-test}, also per electron).

\paragraph{Model.}
PET is a transformer-based graph-neural network~\cite{pozdnyakov2023pet,bigi2026pushing}. 
The densities are predicted mapping the input geometry, atom types, and total charge/spin values to the target coefficients via the PET backbone followed by per-angular-channel nonlinear heads and atom-type-conditioned linear readout layers~\cite{abbott2026unconstrained}.
Equivariance of the coefficients is learned via on-the-fly data augmentation (i.e., each training structure is randomly rotated and/or inverted at every epoch~\cite{pozdnyakov2023pet,domina2026unconstrained}).
The total charge and the spin multiplicity enter as a global embedding added to the features of every atom at each message-passing layer~\cite{bigi2026pushing}.
\PETr{} was pretrained on the coefficients of the total density, fine-tuned on the Coulomb self-energy of its error, and finally trained jointly on the total and spin densities.
At inference, the coefficients are shifted by the smallest change, in the Coulomb metric, that restores the number of electrons of the density and the number of unpaired electrons of the spin density.
It was trained with \texttt{metatrain}~\cite{bigi2026metatomic} on 64 GH200 GPUs in about 11,000 GPU-hours; architecture, hyperparameters and training stages are given in SI, Sec.~\ref{sec:si-models}.
\paragraph{Error estimates.}
The weight deviations of each block of element $Z$ and angular momentum $\lambda$ are Gaussian, with covariance equal to the inverse of the regularized covariance of the last-layer features of the training atoms of element $Z$.
Each member is normalized, and the members are recentered on the prediction.
The factors are fitted so that, over the validation structures, the predicted Coulomb self-energy of the error matches the observed one in the median, by least squares on the logarithm of their ratio, with a weak penalty that draws the factors of each element towards a common value.
Details, the gap head and the gap correction are given in SI, Sec.~\ref{sec:si-uq}.

\paragraph{Reference calculations.}
DFT references outside OMol25 are $\omega$B97M-V/def2-TZVPD calculations with density fitting in PySCF and GPU4PySCF~\cite{sun2020pyscf,li2025gpu4pyscf} (ORCA~\cite{neese2025orca6} for the KSI clusters), unrestricted for open-shell systems.
The ESP and fields of a predicted density were computed from the coefficients with analytic Coulomb integrals of the auxiliary functions.
Unless stated otherwise, geometries were built from SMILES with RDKit (ETKDG~\cite{riniker2015etkdg} and MMFF94~\cite{halgren1996mmff94}) and relaxed with the PET-OMol-L interatomic potential~\cite{bigi2026pushing}.

\paragraph{Observables.}
In KSI, variants were built from PDB 5KP4~\cite{wu2016ksi} with 19-nortestosterone and sampled at 300~K by ML/MM dynamics with mechanical embedding in OpenMM.
In Y16S, two waters were placed in the cavity left by Tyr16, as in a crystal structure of the Y16S/D40N mutant~\cite{kraut2010ksi}.
The rest of the protein, the solvent and the interaction between the two regions were described by ff14SB~\cite{maier2015ff14sb} and TIP3P~\cite{jorgensen1983tip3p}, with hydrogen link atoms at the peptide bonds cut by the ML region.
The field on the C=O bond was computed from the density predicted for the whole periodic cell except the ligand, normalized to the total charge of the cell.
The measured C=O frequencies were converted into fields with the quadratic calibration of the vibrational Stark experiment~\cite{fried2014ksi}.
The $\sigma$-hole is the maximum of the ESP on the 0.001~a.u.\ isosurface of the predicted density on the cap of iodine.
For hydrogen-bond basicity, $V_\mathrm{min}$ is the minimum of the ESP within 2~\AA{} of the acceptor atom, and p$K_\mathrm{BHX}$ values were taken from a public compilation of measurements~\cite{kenny2020pkbhxdata}.
The share of complexes at each site of a bifunctional base follows from the p$K_\mathrm{BHX}$ that each $V_\mathrm{min}$ predicts through the linear fit.
Core-electron binding energies were estimated from the ESP at the nucleus of the neutral molecule and of its equivalent-core cation~\cite{jolly1970equivalent}, averaged over the two~\cite{davis1972relaxation}.
One offset per element was fitted over its sites, and the binding energies were compared with gas-phase measurements~\cite{jolly1984core,golze2020core} (SI, Sec.~\ref{sec:si-xps}).

\paragraph{Density maps.}
For the maps of Fig.~\ref{fig:spin} and SI Fig.~\ref{fig:si-spin-maps}, the organic radicals were built from SMILES and relaxed as doublets with PET-OMol-L, and Fe(H$_2$O)$_6^{2+}$ was built with ideal octahedral angles.
The dibromoethane cluster and the acetaldehyde--formic acid cation are OMol25 test structures, with their OMol25 density matrices as references.
The TCNE--benzoquinone anion was started from a guess with the electron on TCNE, since the default guess converges to a state 1.5~eV higher.
Atomic populations are integrals over Becke cells~\cite{becke1988multicenter}.
Molecular renders were made with UCSF ChimeraX~\cite{meng2023chimerax}.

\section*{Data Availability Statement}

All data supporting this publication, as well as training, analysis, and plotting scripts, are available in the Materials Cloud archive associated with this publication \textit{[Reference to be included upon publication]}. The OMol25 dataset is available from the respective developers.  
\begin{acknowledgments}
P.P. acknowledges support from Novartis Pharma AG through the Early Impact Fund (EIF),
agreement No.~6480068733. J.W.A. and M.C. acknowledge support from the European Research
Council (ERC) under the European Union's Horizon 2020 research and innovation programme
(grant agreement No.~101001890, FIAMMA). M.C. acknowledges support from the Swiss National
Science Foundation (SNSF), project ID 200020\_214879. The calculations were run at the Swiss
National Supercomputing Centre (CSCS), in part through the Swiss AI Initiative, under project
IDs lp161, lp140 and ab010, and on the EPFL HPC platform (SCITAS).

AI tools have been used to assist with writing plotting and analysis scripts, and to improve the clarity and grammar of the manuscript. \end{acknowledgments}

\clearpage
\onecolumngrid
\setcounter{section}{0}
\setcounter{equation}{0}
\setcounter{figure}{0}
\setcounter{table}{0}
\renewcommand{\thesection}{S\arabic{section}}
\renewcommand{\theequation}{S\arabic{equation}}
\renewcommand{\thefigure}{S\arabic{figure}}
\renewcommand{\thetable}{S\arabic{table}}
 \begin{center}
  {\large\textbf{Supplementary Material}}\\[0.5em]
  \textbf{\papertitle}
\end{center}

\FloatBarrier
\section{Models and training}
\label{sec:si-models}

\paragraph{Earlier models.}
Table~\ref{tab:prior} lists earlier ML models of the electron density with the data they were trained on.
Their reported errors are not listed, because they refer to different densities (the valence density on a grid or the fitted density) and to different test sets.

\begin{table}[!htbp]
\centering
\caption{ML models of the electron density, with their training data, the chemistry it covers, the representation of the predicted density and whether a per-structure uncertainty estimate (UQ) is reported.}
\label{tab:prior}
\small
\begin{tabular}{@{}p{2.9cm}p{4.5cm}p{4.6cm}p{3.1cm}l@{}}
\toprule
Model & Training data (structures) & Chemistry & Representation & UQ \\
\midrule
ML-HK~\cite{brockherde2017bypass} & \raggedright MD snapshots of single molecules (up to 2,000 each; PBE) & \raggedright H$_2$O, benzene, ethane, malonaldehyde; one model per molecule & Fourier basis & no \\
SA-GPR~\cite{fabrizio2019noncovalent} & \raggedright BFDb side-chain dimers (2,291; $\omega$B97X-D/cc-pVQZ) & H, C, N, O; neutral and charged & atom-centered basis & no \\
DeepDFT~\cite{jorgensen2022deepdft} & QM9 (123,835) & H, C, N, O, F; neutral, closed-shell & grid & no \\
OrbNet-Equi~\cite{qiao2022informing} & QM9 & as above$^{a}$ & grid & no \\
ChargE3Net~\cite{koker2024charge3net} & \raggedright QM9 (123,835); Materials Project (106,171) & molecules; inorganic crystals & grid & no \\
eSCN~\cite{fu2024recipe} & QM9 (123,835) & as above & grid & no \\
ELECTRA~\cite{elsborg2025electra} & QM9 (123,835) & as above & floating Gaussians & no \\
e3nn~\cite{rackers2023cracking} & water clusters & H, O; neutral & atom-centered basis & no \\
SALTED~\cite{lewis2021condensed} & \raggedright metals, semiconductors, molecular crystals, ice & \raggedright periodic and molecular; neutral & atom-centered basis & no \\
PET~\cite{abbott2026unconstrained} & \raggedright SCFBench~\cite{liu2026scfbench} (43,862; PBE/def2-SVP) & \raggedright H, C, N, O, F, P, S; up to 20 atoms & atom-centered basis & no \\
EAC-Net~\cite{qin2026eacnet} & Materials Project (48,183) & inorganic crystals & grid & no \\
EdenGNN~\cite{li2025edengnn} & \raggedright Materials Project (51,000; PBE) & \raggedright non-magnetic crystals, 87 elements & \raggedright atom-centered basis (PAW) & no \\
AIDEN~\cite{zeng2026aiden} & \raggedright crystals (140,646; PBE); QM9 (133,885) & \raggedright inorganic crystals, 94 elements; molecules as above & grid & no \\
DenSNet~\cite{wing2026densip} & \raggedright QM7-X (22,500 geometries of 250 molecules; PBE/aug-cc-pVDZ) & \raggedright H, C, N, O, S, Cl; neutral & \raggedright atom-centered Gaussians & no \\
\midrule
\raggedright \PETrho{} (this work) & \raggedright OMol25 (3,986,754; $\omega$B97M-V/def2-TZVPD) & \raggedright 83 elements; charged and open-shell; electron and spin density & atom-centered basis & LLPR \\
\bottomrule
\end{tabular}

\vspace{2pt}
{\footnotesize $^{a}$With features from a semiempirical calculation as input. \PETrho{} is also called \PETr{} in the main text.}
\end{table}

\paragraph{Targets.}
The 4M split of OMol25~\cite{omol} (3,986,754 structures, 218.7 million atoms) was divided as in the Computational Methods with random seed 0.
The coefficients of each atom are grouped into blocks of angular momentum $\lambda$, each holding the $2\lambda + 1$ components of every radial function of that $\lambda$ (from 11 coefficients for H to 200 for the lanthanides).
Before training, a composition baseline was subtracted from the targets and the result was divided by a fixed scale, both computed on the training set and frozen.
The baseline is the mean of each $\lambda = 0$ coefficient over the atoms of each element (zero for the spin density), and the scale is the root mean square of each radial function of each block and element after the baseline is removed, so that all coefficients enter the loss with similar magnitudes.
The spin density of restricted calculations is zero by construction and is excluded from the spin loss.
Unrestricted singlets are kept, with a loss that is invariant to a global change of sign of $m$: their spin density is broken-symmetry, and the two solutions with opposite signs of $m$ are degenerate, so the sign of the reference is arbitrary.

\paragraph{Architecture.}
\PETrho{} is a point-edge transformer (PET)~\cite{pozdnyakov2023pet} as implemented in \texttt{metatrain}~\cite{bigi2026metatomic}.
PET keeps a feature vector (token) on every directed pair of atoms within the cutoff and updates it at each message-passing layer with a transformer over the pairs of the central atom, in which the attention weights are damped by a smooth cutoff function.
The architecture imposes no rotational symmetry, and the model learns the equivariance of the coefficients from training on structures rotated and inverted at random at every step.
We use an adaptive cutoff that keeps about 24 neighbors per atom, within at most 8~\AA, which makes the number of neighbors more uniform than a fixed cutoff, in particular in sparse structures such as separated molecules.
After the last message-passing layer, each block of each target has its own nonlinear head for the atom features and for the pair features, and the pair outputs are summed over neighbors, weighted by the cutoff function.
A linear readout per element maps the head outputs to the coefficients of that block, so that the coefficients of an element are linear in the last-layer features, which the error estimates of Sec.~\ref{sec:si-uq} use.
The total charge $Q$ and spin multiplicity $M$ enter as a learned per-system embedding added to the features of every atom at each message-passing layer~\cite{bigi2026pushing}.
The two targets, the total and the spin density, share the backbone and have separate heads.
Hyperparameters are given in Table~\ref{tab:pet-omol25-hypers}.

\begin{table}[!htbp]
\centering
\small
\setlength{\tabcolsep}{4pt}
\caption{Hyperparameters and training stages of \PETrho{}: adaptive cutoff with about 24 neighbors within 8\,\AA{}, Bump cutoff function (width 0.5\,\AA), 8 attention heads, per-block heads, linear readout per element, charge and spin embedding (charges up to $\pm 10$, multiplicities up to 11), Adam, gradient-norm clip 1.0, learning rate with linear warmup and cosine decay.}
\label{tab:pet-omol25-hypers}
\begin{tabular}{@{}lc@{}}
\toprule
$d_\mathrm{PET}$ / $d_\mathrm{node}$ / $d_\mathrm{ff}$ & 512/1024/1024 \\
Message-passing $\times$ attention layers & $3\times3$ \\
Head width (atom/pair) & 1024/512 \\
Parameters & 145\,M \\
Atoms per batch and GPU & 3000 \\
\midrule
\multicolumn{2}{@{}l}{\emph{1. Pretraining: MSE of the coefficients}} \\
Epochs, peak LR & 300, $7\times10^{-5}$ \\
GH200 GPU-hours & 8,200 \\
\multicolumn{2}{@{}l}{\emph{2. Fine-tuning: Coulomb self-energy of the error, two consecutive runs (peak LR of the first)}} \\
Epochs, peak LR & 70, $7\times10^{-6}$ \\
GH200 GPU-hours & 2,000 \\
\multicolumn{2}{@{}l}{\emph{3. Spin heads on the frozen backbone, unrestricted calculations}} \\
Epochs, peak LR & 30, $1\times10^{-4}$ \\
GH200 GPU-hours & 230 \\
\multicolumn{2}{@{}l}{\emph{4. Joint fine-tuning: total and spin Coulomb losses}} \\
Epochs, peak LR & 17, $1.75\times10^{-6}$ \\
Spin loss weight & 0.045 \\
GH200 GPU-hours & 650 \\
\midrule
Total GH200 GPU-hours & 11,000 \\
\bottomrule
\end{tabular}
\end{table}

\paragraph{Training.}
\PETrho{} was trained in four stages (Table~\ref{tab:pet-omol25-hypers}).
It was first pretrained on the mean squared error of the scaled coefficients of the total density, a loss that is cheap to evaluate and well conditioned.
It was then fine-tuned on the Coulomb self-energy of the error density $\Delta\rho = \rho_\mathrm{ref} - \rho_\mathrm{pred}$,
\begin{equation}
E_\mathrm{H}[\Delta\rho] = \frac{1}{2}\iint \frac{\Delta\rho(\mathbf{r})\,\Delta\rho(\mathbf{r}')}{|\mathbf{r} - \mathbf{r}'|}\,d\mathbf{r}\,d\mathbf{r}',
\end{equation}
which is the metric in which the targets were fitted.
Spin heads were trained next on the frozen backbone, so that they could not degrade the total density, on the unrestricted calculations only.
Finally, all parameters were fine-tuned on the weighted sum of the Coulomb losses of the total and spin densities.
For each stage, the exported model is the one with the lowest validation loss.
Every stage ran on 16 nodes with four GH200 GPUs each, on the Alps system at CSCS.
The cost in Table~\ref{tab:pet-omol25-hypers} is the time per epoch of each stage with the current version of \texttt{metatrain}, measured on the same 64 GPUs, times the number of epochs: about 11,000 GPU-hours, of which pretraining takes about three quarters.

\paragraph{Inference.}
All results use the normalization of the Computational Methods unless stated otherwise.
For the spin density it also removes the spurious spin of closed shells: on the closed-shell singlet of Fe(H$_2$O)$_6^{2+}$ the raw spin head integrates to 0.44~e ($\int|m| = 0.67$~e) instead of zero, so that a spin map of a closed-shell structure vanishes only once it is constrained to $2S = 0$.
The DFT timings of Fig.~\ref{fig:accuracy}c of the main text are $\omega$B97M-V/def2-TZVPD single points in GPU4PySCF with density fitting, default grids and convergence thresholds and the minao initial guess, on the neutral closed-shell structures.
They are timed from the construction of the molecule to the convergence of the SCF (means of three repeats for runs under 10~min, one run above).
The potential on $10^4$ points takes longer than the prediction at every size.

\FloatBarrier
\section{Test-set evaluation}
\label{sec:si-test}

\paragraph{Error measures.}
The density error of a structure is $\int |\rho_\text{pred} - \rho_\text{ref}|\,d\mathbf{r}$, evaluated on a PySCF integration grid (level 3) against two references: the DFT density from the density matrix, and the fitted density that the model learns.
The error of the fitted density against DFT is the `RI floor' of each structure.
The main text reports this error per atom; here we also divide it by the number of electrons $N_\text{el}$, which gives the normalized mean absolute error (NMAE) used in earlier work on molecular densities.
Per electron, the many core electrons of heavy atoms, which are easy to predict, dilute the error, so per atom the structures with heavy elements are compared on the same footing as organic ones.
Over the 39,845 test structures, the median error of \PETrho{} against DFT is 41~me per atom and 1.01\% per electron, against 40~me per atom and 0.97\% for the fitted density.
The spin-density NMAE is $\int|m_\text{pred} - m_\text{ref}|\,d\mathbf{r} / \int|m_\text{ref}|\,d\mathbf{r}$, normalized by the total unsigned spin of the reference instead of $N_\text{el}$, since $m$ integrates to $2S$ and not to the number of electrons.
The 20 noble-gas test structures (clusters of noble-gas atoms with water) are reported separately: the fit differs from DFT by an NMAE of 1.26\% and \PETrho{} by 1.27\% (medians); the statistics below cover the other 39,845 test structures.
On every test structure, the number of electrons integrated on the grid is within 0.012~e of the exact one.
Table~\ref{tab:nmae-category} gives the NMAE by subset and source.

\begin{table}[!htbp]
\centering
\footnotesize
\caption{NMAE (\%) by source of the OMol25 structures, in the subsets of Fig.~\ref{fig:accuracy}b of the main text: median and 90th percentile (P90) against DFT of the fitted density (Fit) and of \PETrho{} (L), and median of \PETrho{} against the fitted density.}
\label{tab:nmae-category}
\begin{tabular*}{\linewidth}{@{\extracolsep{\fill}}llrrrrrr@{}}
\toprule
 & & & \multicolumn{4}{c}{Against DFT} & Against Fit \\
\cmidrule(lr){4-7}\cmidrule(l){8-8}
Subset & Source & $n$ & Fit & Fit P90 & L & L P90 & L \\
\midrule
small organics & Orbnet Denali & 511 & 0.92 & 1.01 & 0.92 & 1.01 & 0.11 \\
 & SPICE & 444 & 0.95 & 1.20 & 0.96 & 1.21 & 0.10 \\
 & GEOM & 1998 & 0.97 & 1.07 & 0.97 & 1.08 & 0.10 \\
 & torsion profiles & 42 & 1.00 & 1.04 & 1.00 & 1.05 & 0.09 \\
 & ANI-2x & 2156 & 1.00 & 1.19 & 1.02 & 1.20 & 0.15 \\
 & ANI-1xbb & 7372 & 0.97 & 1.20 & 1.02 & 1.30 & 0.22 \\
\midrule
organic reactions & reaction mechanisms & 526 & 0.91 & 1.02 & 0.93 & 1.07 & 0.14 \\
 & Transition1x & 2137 & 1.08 & 1.20 & 1.10 & 1.22 & 0.17 \\
 & open-shell RGD1 & 753 & 1.16 & 1.28 & 1.20 & 1.35 & 0.22 \\
\midrule
biomolecules & nucleic acids & 649 & 0.92 & 0.99 & 0.92 & 0.99 & 0.10 \\
 & PDB pockets & 201 & 0.97 & 1.02 & 0.97 & 1.02 & 0.11 \\
 & PDB fragments & 1860 & 0.98 & 1.04 & 0.99 & 1.04 & 0.10 \\
 & protein cores & 354 & 1.02 & 1.07 & 1.03 & 1.07 & 0.08 \\
 & protein interfaces & 450 & 1.03 & 1.07 & 1.03 & 1.07 & 0.09 \\
 & ML protein interfaces & 619 & 1.02 & 1.07 & 1.03 & 1.14 & 0.09 \\
 & solvated protein & 3823 & 1.27 & 1.47 & 1.30 & 1.56 & 0.12 \\
\midrule
electrolytes & electrolytes & 4976 & 0.96 & 1.16 & 0.99 & 1.19 & 0.10 \\
 & electrolyte clusters & 142 & 1.05 & 1.23 & 1.06 & 1.24 & 0.11 \\
 & ML electrolytes & 516 & 1.05 & 1.22 & 1.07 & 1.24 & 0.10 \\
 & path-integral MD & 187 & 1.09 & 1.48 & 1.09 & 1.47 & 0.12 \\
 & droplets & 187 & 1.20 & 1.46 & 1.21 & 1.46 & 0.12 \\
\midrule
scaled separations & electrolyte separations & 308 & 0.94 & 1.16 & 1.04 & 1.35 & 0.28 \\
 & scaled separations & 532 & 1.07 & 1.31 & 1.13 & 1.45 & 0.10 \\
\midrule
metal complexes & metal complexes & 5374 & 0.84 & 0.94 & 0.88 & 1.03 & 0.18 \\
 & recomputed & 76 & 0.85 & 0.96 & 0.89 & 1.07 & 0.20 \\
 & ML metal complexes & 776 & 0.85 & 0.96 & 0.92 & 1.21 & 0.21 \\
 & low-spin & 129 & 0.84 & 0.96 & 0.94 & 1.11 & 0.30 \\
 & metal hydrides & 34 & 0.85 & 0.96 & 1.02 & 1.28 & 0.46 \\
\midrule
metal reactivity & metal reactivity & 1582 & 0.94 & 1.08 & 1.15 & 1.62 & 0.50 \\
\midrule
electrolyte redox, & electrolyte reactivity & 670 & 0.81 & 1.04 & 1.16 & 2.45 & 0.66 \\
\quad reactivity & electrolyte redox & 461 & 1.01 & 1.33 & 1.27 & 1.83 & 0.46 \\
\bottomrule
\end{tabular*}
\end{table}

\paragraph{Training against test error.}
We evaluated \PETrho{} against the fitted density on 150 random training structures per subset (60 open-shell structures per subset for the spin density) and compared the medians with those of the test set.
Each subset mixes sources with different errors, so the training structures were weighted to give each source the same share as among the test structures of its subset (Table~\ref{tab:train-test}).
The training error of the density is about 60\% of the test error for metal reactivity and for electrolyte redox and reactivity, with a larger gap for open-shell (51--58\%) than for closed-shell structures (81\%), and agrees with the test error within 10\% elsewhere.
The spin-density error of \PETrho{} on training structures is 71--108\% of the test error.

\begin{table}[!htbp]
\centering
\caption{Median error against the fitted density on training (weighted) and test structures: density NMAE (\%) and, for open-shell structures, spin-density NMAE (\%) of \PETrho{}.}
\label{tab:train-test}
\begin{tabular*}{\linewidth}{@{\extracolsep{\fill}}lcccc@{}}
\toprule
 & \multicolumn{2}{c}{$\rho$} & \multicolumn{2}{c}{$m$} \\
\cmidrule(lr){2-3}\cmidrule(l){4-5}
Subset & train & test & train & test \\
\midrule
small organics & 0.18 & 0.18 & 37 & 43 \\
organic reactions & 0.17 & 0.17 & 47 & 55 \\
biomolecules & 0.10 & 0.10 & 43 & 45 \\
electrolytes & 0.11 & 0.11 & 27 & 33 \\
scaled separations & 0.17 & 0.19 & 47 & 67 \\
metal complexes & 0.17 & 0.19 & 58 & 53 \\
metal reactivity & 0.32 & 0.50 & 45 & 62 \\
electrolyte redox, reactivity & 0.31 & 0.57 & 57 & 71 \\
\bottomrule
\end{tabular*}
\end{table}

\paragraph{Several SCF solutions.}
A training label that the model reproduces but cannot predict for a new structure would explain part of the gap between training and test error, so we checked whether the reference of slowly converging open shells is unique.
We selected 30 open-shell test structures of metal reactivity, electrolytes, biomolecules and organic reactions whose reference SCF took more than 150 cycles, and 10 open-shell controls that took at most 30.
For each, we ran unrestricted $\omega$B97M-V/def2-TZVPD calculations in GPU4PySCF from five initial guesses (the reference density matrix, MINAO, the superposition of atomic densities, extended H\"uckel and the core Hamiltonian), with at most 150 or 300 cycles, and merged converged solutions within $10^{-5}$~$E_\text{h}$ in energy and $10^{-5}$ in Coulomb distance $\Delta\mathbf{c}^\top \mathbf{J}\,\Delta\mathbf{c}$, where $\Delta\mathbf{c}$ is the difference of their fitted coefficients and $\mathbf{J}$ the Coulomb metric of the auxiliary basis.
The controls converge from at least three guesses to the reference solution, and \PETrho{} lies on it.
Of the 30 slow structures, 8 reach no converged solution within the cycle limit.
Of the other 22, 16 have two or more solutions, and 8 have a solution more than 1~m$E_\text{h}$ below the reference, by up to 48~m$E_\text{h}$.
\PETrho{} is closer to another solution than to the reference for 10 of the 22 structures, but its distance to the closest solution remains 4 to 200 times the median per-atom error on training open-shell structures.
For the 16 structures with two or more solutions, the distance of \PETrho{} from the reference is comparable to the largest distance between two solutions (median ratio 1.0, from 0.1 to 1.4).
The training open-shell structures whose SCF took more than 150 cycles have the same median per-atom error as the others, so their labels are fitted as well.

\FloatBarrier
\section{Electrostatic potential}
\label{sec:si-esp}

The median root-mean-square error (RMSE) of the surface ESP of \PETrho{} is 2.9 times that of the fitted density in metal reactivity, 4.3 times in electrolyte reactivity and 8.3 times in electrolyte redox, against 1.0--1.4 times in the subsets of small organics, biomolecules and electrolytes (Table~\ref{tab:esp}).
The ESP was evaluated on 1,500 random points of the $\rho = 0.001$~a.u.\ isosurface of the DFT density for all 39,845 test structures.
The DFT potential was computed exactly from the density matrix with GPU4PySCF, and the potentials of the fitted and predicted densities with libcint in double precision.
Over all structures, the median RMSE against DFT is 0.83~kcal\,mol$^{-1}$\,e$^{-1}$ for the fitted density and 1.04~kcal\,mol$^{-1}$\,e$^{-1}$ for \PETrho{}.

\begin{table}[!htbp]
\centering
\footnotesize
\caption{Median RMSE of the electrostatic potential on the molecular surface against DFT (kcal\,mol$^{-1}$\,e$^{-1}$) by OMol25 subset and source, for the fitted density (Fit) and \PETrho{} (L).}
\label{tab:esp}
\begin{tabular*}{\linewidth}{@{\extracolsep{\fill}}llrrr@{}}
\toprule
Subset & Source & $n$ & Fit & L \\
\midrule
small organics & Orbnet Denali & 511 & 0.76 & 0.85 \\
 & SPICE & 444 & 0.81 & 0.89 \\
 & GEOM & 1998 & 0.83 & 0.95 \\
 & torsion profiles & 42 & 0.89 & 0.96 \\
 & ANI-2x & 2156 & 0.66 & 0.80 \\
 & ANI-1xbb & 7372 & 0.73 & 1.05 \\
\midrule
organic reactions & reaction mechanisms & 526 & 0.80 & 1.08 \\
 & Transition1x & 2137 & 0.83 & 0.94 \\
 & open-shell RGD1 & 753 & 0.96 & 1.13 \\
\midrule
biomolecules & nucleic acids & 649 & 0.65 & 0.82 \\
 & PDB pockets & 201 & 0.91 & 1.07 \\
 & PDB fragments & 1860 & 0.91 & 1.01 \\
 & protein cores & 354 & 0.97 & 1.02 \\
 & protein interfaces & 450 & 0.96 & 1.01 \\
 & ML protein interfaces & 619 & 0.96 & 1.05 \\
 & solvated protein & 3823 & 0.98 & 1.07 \\
\midrule
electrolytes & electrolytes & 4976 & 0.82 & 0.99 \\
 & electrolyte clusters & 142 & 0.96 & 1.10 \\
 & ML electrolytes & 516 & 0.97 & 1.06 \\
 & path-integral MD & 187 & 0.87 & 1.00 \\
 & droplets & 187 & 0.98 & 1.05 \\
\midrule
scaled separations & electrolyte separations & 308 & 0.81 & 1.59 \\
 & scaled separations & 532 & 0.90 & 1.18 \\
\midrule
metal complexes & metal complexes & 5374 & 0.81 & 1.20 \\
 & recomputed & 76 & 0.76 & 1.30 \\
 & ML metal complexes & 776 & 0.78 & 1.46 \\
 & low-spin & 129 & 0.81 & 1.70 \\
 & metal hydrides & 34 & 0.82 & 3.33 \\
\midrule
metal reactivity & metal reactivity & 1582 & 0.89 & 2.61 \\
\midrule
electrolyte redox, & electrolyte reactivity & 670 & 0.58 & 2.50 \\
\quad reactivity & electrolyte redox & 461 & 0.91 & 7.56 \\
\bottomrule
\end{tabular*}
\end{table}

\FloatBarrier
\section{Ketosteroid isomerase}
\label{sec:si-ksi}

\paragraph{Preparation and equilibration.}
Variants were built from PDB 5KP4 with 19-nortestosterone (19-NT).
In Y16S, two waters were added in the cavity left by the Tyr16 side chain, one at the position of its hydroxyl O and one 2.75~\AA{} from it towards the former ring, as in the crystal structure of Y16S/D40N~\cite{kraut2010ksi}.
Hydrogens were added with OpenMM at pH 7.0, with Asp103 neutral in both chains except in the variant with deprotonated Asp103, and the hydroxyl hydrogens of the residues that donate hydrogen bonds to the ligand were turned towards its carbonyl.
The protein was described with ff14SB~\cite{maier2015ff14sb}, the ligand with OpenFF Sage 2.2.1 and AM1-BCC charges, and the system was solvated in TIP3P water~\cite{jorgensen1983tip3p} with 1~nm of padding and 0.15~M NaCl, with hydrogen mass repartitioning.
After minimization, the system was heated from 30 to 300~K in 200~ps at constant volume with the heavy atoms of the protein and the ligand restrained to their positions (1000~kJ\,mol$^{-1}$\,nm$^{-2}$), and equilibrated at constant pressure for 250~ps at each of the restraint constants 1000, 100, 10 and 0~kJ\,mol$^{-1}$\,nm$^{-2}$.
A 10~ns constant-pressure run with a 4~fs time step followed, with flat-bottom restraints that keep the ligand O between 2.5 and 3.5~\AA{} from the Tyr16 hydroxyl O, between 3.5 and 4.5~\AA{} from the Asp40 carboxylate O and between 5.5 and 6.5~\AA{} from the Asp103 C$_\alpha$, and the 12 starting frames of the ML/MM runs were selected from it for a crystal-like hydrogen-bond geometry.
\paragraph{ML/MM dynamics.}
Production runs used ML/MM dynamics in OpenMM with mechanical embedding: the residues with a heavy atom within 10~\AA{} of 19-NT in the starting frame are described by \texttt{PET-OMol-L}~\cite{bigi2026pushing} as a machine-learned potential, and the rest by the force field, which also gives the interaction between the two regions.
Bonds that cross the boundary were cut at the peptide bonds and capped with hydrogen link atoms, the force-field terms inside the region were removed, and the potential was evaluated on the region and its caps without periodic images.
Each variant was run as 12 trajectories of 100~ps at 300~K, with a Langevin thermostat (friction 1~ps$^{-1}$) and a 2~fs time step, without restraints.
In every variant, a few protons of surface salt bridges of chain B, 10--16~\AA{} from the ligand, transfer within the ML region, likely an artifact of mechanical embedding, in which the solvent does not screen the charges of the ML region.
Counted as unit charges, they change the shifts from the wild type (WT) by at most 2~MV\,cm$^{-1}$, and by 7~MV\,cm$^{-1}$ in Y16F with Asp103 ionized.
The radius of the region was chosen by convergence: the field of WT changes by 2~MV\,cm$^{-1}$ between 8 and 10~\AA{}, against 8~MV\,cm$^{-1}$ between 6 and 10~\AA{}.

\paragraph{Field.}
The field on the C=O was computed every 1~ps from the density predicted for the whole periodic cell without the ligand (protein, waters and ions, about 40,000 atoms), and averaged from 20 to 100~ps of each trajectory.
The electron count of the cell is fixed by a single normalization, which shifts only the $s$-function coefficients, because the Coulomb metric does not scale to the cell, and changes the field by less than 0.2~MV\,cm$^{-1}$.
Each prediction takes about 10~s on one GH200 GPU; one frame of 7,680 (WT, seed 4, 63~ps) exceeded the GPU memory and was skipped.
As a check, the field was also computed from the density of each chain predicted separately, with the waters whose O is within 6~\AA{} of any ligand atom added to chain B, and from the remaining waters and ions as point charges (partitioned field).
For the neutral variants, the partitioned field is within 3.5~MV\,cm$^{-1}$ of that of the whole cell, and the shifts from WT agree within 3~MV\,cm$^{-1}$.
With Asp103 ionized, the partitioned field is 5~MV\,cm$^{-1}$ stronger, because Na$^+$ ions enter the active site and act as point charges on a density predicted without them.
The decompositions into residue terms below use the partitioned field.
The field is projected on the unit vector from C to O, along which it is negative in the enzyme, and the shift of a variant is its field minus that of WT, so that a positive shift is a weaker field.
Fields were compared with the experimental frequencies converted with the quadratic calibration of the vibrational Stark experiment~\cite{fried2014ksi} (Table~\ref{tab:ksi-fields}).
Over the variants of Fig.~\ref{fig:observables}a of the main text (WT, D103N, D103L, Y16S and Y16F), the mean absolute difference between the computed field, the median over the 12 seeds, and the experimental field is 9.6~MV\,cm$^{-1}$ for the density and 12.5~MV\,cm$^{-1}$ for the point charges (9.3 and 11.7~MV\,cm$^{-1}$ with D40N), and 4.4 and 6.8~MV\,cm$^{-1}$ without Y16F.
The experimental fields of these variants span 71~MV\,cm$^{-1}$, from $-129$ (WT) to $-58$~MV\,cm$^{-1}$ (Y16F).
The DFT points of Fig.~\ref{fig:observables}a of the main text are the field of \PETrho{} of each variant plus the mean of DFT minus \PETrho{} over the active-site clusters of that variant, described below.

\begin{table}[!htbp]
\centering
\small
\caption{Field on the C=O bond (MV\,cm$^{-1}$): experiment and median over 12 seeds $\pm$ standard error, and in parentheses the shift relative to WT (positive is a weaker field).}
\label{tab:ksi-fields}
\begin{tabular*}{\linewidth}{@{\extracolsep{\fill}}lccc@{}}
\toprule
Variant & Experiment & Density & Point charges \\
\midrule
WT & $-129.0$ & $-135.5 \pm 1.6$ & $-139.8 \pm 1.2$ \\
D103N & $-121.5$ (7.5) & $-120.1 \pm 1.3$ (15.4) & $-123.8 \pm 1.2$ (16.0) \\
D40N & $-122.2$ (6.8) & $-114.2 \pm 0.9$ (21.2) & $-114.4 \pm 0.8$ (25.4) \\
Y16S & $-90.3$ (38.7) & $-97.2 \pm 2.5$ (38.2) & $-102.3 \pm 2.4$ (37.4) \\
Y16S, no cavity water & & $-86.5 \pm 1.9$ (49.0) & $-88.1 \pm 2.0$ (51.7) \\
D103L & $-86.6$ (42.4) & $-89.4 \pm 3.5$ (46.1) & $-88.6 \pm 3.4$ (51.2) \\
Y16F & $-57.8$ (71.2) & $-88.1 \pm 2.7$ (47.4) & $-93.0 \pm 2.7$ (46.8) \\
\bottomrule
\end{tabular*}
\end{table}

\paragraph{Density against DFT.}
To separate the error of the density from that of the sampling, we compared the field of \PETrho{} with DFT ($\omega$B97M-V/def2-TZVPD, ORCA 6.1.1~\cite{neese2025orca6}, RIJCOSX) on pieces of the active site cut from the trajectories.
On isolated residues and waters, the two agree within 0.4~MV\,cm$^{-1}$ for charged residues and waters, and within 7\% for the fragment that contains Tyr16.
The 36 active-site clusters (53--72 atoms) were cut from one frame of each of six seeds per variant of earlier runs with a 6~\AA{} region (WT, D103N, D103L, Y16S without cavity water, Y16F, and Y16F with Asp103 deprotonated).
They contain the side chains of residues 16, 32, 40, 57 and 103 of chain B (cut at C$_\alpha$--C$_\beta$ and capped with H) and the waters with O within 4~\AA{} of the ligand O, without the ligand.
The point charges are those of the force field for the same atoms, with each cap H carrying the remaining charge of its residue.
Against DFT, the mean absolute error of the total field is 1.6~MV\,cm$^{-1}$ (maximum 8.7) for the density and 3.8~MV\,cm$^{-1}$ (maximum 12.7) for the point charges, and 2.2 and 3.7~MV\,cm$^{-1}$ for the shift from WT.
Per variant, the mean absolute error of the density is at most 3.0~MV\,cm$^{-1}$, and the mean difference that corrects the field in Fig.~\ref{fig:observables}a of the main text at most 2.4~MV\,cm$^{-1}$.

Point charges lack by construction the part of the field that comes from the mutual polarization of fragments.
With the density, the cooperative part of the Tyr16 and Asp103 terms in the clusters (their field inside the cluster minus their field alone) is within 0.9~MV\,cm$^{-1}$ of DFT in every variant.

\paragraph{Tyr16 mutants.}
In Y16S, the water at the position of the Tyr16 hydroxyl is within 3.5~\AA{} of the ligand O for 70--90\% of the run in 8 of the 12 seeds, and for less than half in the other 4.
With the cavity waters, the field of the density is 11~MV\,cm$^{-1}$ stronger than without them, and the two are within 7 and 4~MV\,cm$^{-1}$ of experiment (Table~\ref{tab:ksi-fields}).
Y16F deviates most from the rate--field line in Fig.~\ref{fig:observables}a of the main text: its field is too strong by 30~MV\,cm$^{-1}$ for the density and 35~MV\,cm$^{-1}$ for the point charges, and its shift from WT, a weakening of the field, is about 24~MV\,cm$^{-1}$ smaller than measured with both.
Since the cluster fields of Y16F are within 2~MV\,cm$^{-1}$ of DFT on average, this deviation comes from the sampled structures and not from the density.
A larger ML region does not remove it: with 6~\AA{}, the field of Y16F is 9.5~MV\,cm$^{-1}$ weaker and that of WT 8~MV\,cm$^{-1}$.
The experiment shows the same anomaly: point charges on the crystal structure give 30~MV\,cm$^{-1}$ for the Tyr16 hydroxyl, against the 84~MV\,cm$^{-1}$ that its removal costs in the measurement~\cite{fried2014ksi}.

\paragraph{Consistency checks for Y16F.}
Harmonic normal modes of 19-NT were computed with \texttt{PET-OMol-L} in clusters of the residues and waters within 7~\AA{} of the ligand O, on one frame of each seed.
In WT, D103N, D103L and Y16S, the C=O stretch follows the field computed on the same frame with a slope of 0.78~cm$^{-1}$ per MV\,cm$^{-1}$ ($r = 0.75$), close to that of the experimental calibration, and the residual of Y16F from this line ($+6 \pm 4$~cm$^{-1}$) is comparable to those of these variants.
Starting Y16F from the crystal structure of apo Y16F (PDB 1EA2), or with and without a water hydrogen-bonded to the C=O, changes its median field by at most 4~MV\,cm$^{-1}$.
With Asp103 ionized, the field of Y16F is $-45$~MV\,cm$^{-1}$, 13~MV\,cm$^{-1}$ weaker than measured.
This state would require the p$K_\mathrm{a}$ of Asp103 to fall by more than two units from WT; that of the equivalent Asp99 in the free KSI of \emph{C.~testosteroni} is above 8.5~\cite{thornburg1998asp99}.
The protonation state of Asp103 in Y16F has not been measured, and a free-energy calculation of its p$K_\mathrm{a}$ in the enzyme~\cite{xu2026acidbase,kabylda2026so3lr} would show which of the two states the enzyme adopts.

\FloatBarrier
\section{$\sigma$-hole of iodobenzenes}
\label{sec:si-sigma}

Geometries were built from SMILES with RDKit (ETKDG and MMFF) and relaxed with \texttt{PET-OMol-L} to a maximum force of $10^{-3}$~eV\,\AA$^{-1}$.
The $\rho = 0.001$~a.u.\ isosurface of the predicted density was obtained by marching cubes on a grid of 0.15~\AA, and each vertex was moved onto the exact isosurface by one Newton step along the gradient of the density.
The ESP of the predicted density was evaluated on these points.
The $\sigma$-hole maximum $V_{S,\max}$ is the largest ESP on the cap of iodine, the points within 2.6~\AA{} of the iodine nucleus and within 30$^\circ$ of the C--I axis.
The $\sigma_p$ values are from a standard compilation~\cite{hansch1991survey}.
For all 20 molecules, the ESP of the $\omega$B97M-V/def2-TZVPD density and of its Coulomb fit on \texttt{def2-universal-jfit} was evaluated on the same points.
$V_{S,\max}$ of \PETrho{} is higher than that of the DFT density by 0.81~kcal\,mol$^{-1}$ on average and by 0.41 to 1.18~kcal\,mol$^{-1}$ (Table~\ref{tab:sigma-hole}).
About 0.5~kcal\,mol$^{-1}$ of this difference is already in the fitted density, which lies 0.46 to 0.54~kcal\,mol$^{-1}$ above the DFT density for every molecule, so that \PETrho{} differs from the fit by 0.35~kcal\,mol$^{-1}$ on average and by at most 0.71~kcal\,mol$^{-1}$, for X = NHMe.
Against the fit, \PETrho{} is higher by 0.3 to 0.7~kcal\,mol$^{-1}$ for the donors and alkyl groups and within 0.3~kcal\,mol$^{-1}$ for the halogens and the strongest acceptors.
Against $\sigma_p$, the slope is $8.4 \pm 0.6$~kcal\,mol$^{-1}$ for \PETrho{} and $8.8 \pm 0.6$~kcal\,mol$^{-1}$ for DFT and for its fit, with $r = 0.96$ for all three.

\begin{table}[!htbp]
\centering
\small
\caption{$\sigma$-hole maximum $V_{S,\max}$ (kcal\,mol$^{-1}$) of \textit{para}-X-iodobenzenes from \PETrho{}, from the DFT density and from its fit on \texttt{def2-universal-jfit}, with the Hammett constant $\sigma_p$~\cite{hansch1991survey}.}
\label{tab:sigma-hole}
\begin{tabular}{@{}lcccc@{\hspace{2em}}lcccc@{}}
\toprule
X & $\sigma_p$ & \PETrho{} & DFT & Fit & X & $\sigma_p$ & \PETrho{} & DFT & Fit \\
\midrule
NMe$_2$ & $-0.83$ & 15.54 & 14.45 & 14.92 & F & 0.06 & 22.71 & 21.94 & 22.45 \\
NHMe & $-0.70$ & 16.00 & 14.82 & 15.29 & I & 0.18 & 23.37 & 22.95 & 23.47 \\
NH$_2$ & $-0.66$ & 16.70 & 15.72 & 16.18 & Cl & 0.23 & 23.50 & 22.88 & 23.38 \\
OH & $-0.37$ & 19.20 & 18.43 & 18.92 & Br & 0.23 & 23.81 & 23.13 & 23.66 \\
OMe & $-0.27$ & 18.33 & 17.46 & 17.95 & CHO & 0.42 & 25.87 & 25.16 & 25.68 \\
Me & $-0.17$ & 18.87 & 17.83 & 18.30 & COOMe & 0.45 & 23.44 & 22.63 & 23.13 \\
Et & $-0.15$ & 18.88 & 17.82 & 18.29 & COMe & 0.50 & 24.31 & 23.40 & 23.91 \\
H & 0.00 & 20.07 & 19.10 & 19.58 & CF$_3$ & 0.54 & 26.05 & 25.53 & 26.04 \\
SMe & 0.00 & 21.76 & 20.78 & 21.28 & CN & 0.66 & 28.65 & 28.21 & 28.74 \\
NHCOMe & 0.00 & 19.64 & 18.59 & 19.08 & NO$_2$ & 0.78 & 29.52 & 29.11 & 29.65 \\
\bottomrule
\end{tabular}
\end{table}

\FloatBarrier
\section{Hydrogen-bond basicity}
\label{sec:si-hbond}

\paragraph{Data and geometries.}
We selected 44 neutral acceptors in seven common families (10 carbonyls, 6 ethers and alcohols, 7 pyridines and azoles, 5 nitriles, 5 amines, 5 sulfoxides, 6 phosphoryl compounds) from a public compilation of measured p$K_\mathrm{BHX}$ values~\cite{kenny2020pkbhxdata}, which gives the primary reference of each measurement against 4-fluorophenol in CCl$_4$.
Table~\ref{tab:pkbhx-data} lists the acceptors with their measured and computed values.
Geometries were generated with RDKit (ETKDG, lowest of 20 MMFF conformers) and relaxed with \texttt{PET-OMol-L} to a maximum force of $10^{-3}$~eV\,\AA$^{-1}$.

\begin{table}[!htbp]
\centering
\footnotesize
\caption{Hydrogen-bond acceptors: measured p$K_\mathrm{BHX}$ from a public compilation~\cite{kenny2020pkbhxdata}, which gives the primary reference of each value, and $V_\mathrm{min}$ (kcal\,mol$^{-1}$) of \PETrho{} and of the full DFT density.}
\label{tab:pkbhx-data}
\begin{tabular}{@{}llrrr@{}}
\toprule
Family & Acceptor & p$K_\mathrm{BHX}$ & $V_\mathrm{min}$ \PETrho{} & $V_\mathrm{min}$ DFT \\
\midrule
carbonyl & acetone & $1.18$ & $-48.2$ & $-48.7$ \\
 & cyclohexanone & $1.39$ & $-49.8$ & $-50.6$ \\
 & acetaldehyde & $0.65$ & $-44.8$ & $-44.8$ \\
 & benzaldehyde & $0.78$ & $-45.8$ & $-45.6$ \\
 & 1,1,1-trifluoropropan-2-one & $-0.06$ & $-33.6$ & $-33.7$ \\
 & \textit{N},\textit{N}-dimethylformamide & $2.10$ & $-57.0$ & $-57.2$ \\
 & \textit{N},\textit{N}-dimethylacetamide & $2.44$ & $-58.8$ & $-59.3$ \\
 & 1-methylpyrrolidin-2-one & $2.38$ & $-59.4$ & $-59.7$ \\
 & methyl acetate & $1.00$ & $-45.9$ & $-46.4$ \\
 & ethyl acetate & $1.07$ & $-46.3$ & $-46.9$ \\
\midrule
ether, alcohol & tetrahydrofuran & $1.28$ & $-49.0$ & $-49.4$ \\
 & ethoxyethane & $1.01$ & $-45.9$ & $-46.7$ \\
 & oxetane & $1.36$ & $-48.9$ & $-49.3$ \\
 & methanol & $0.82$ & $-46.2$ & $-46.0$ \\
 & ethanol & $1.02$ & $-47.0$ & $-47.1$ \\
 & 2,2,2-trifluoroethanol & $-0.28$ & $-33.4$ & $-33.0$ \\
\midrule
pyridine, azole & pyridine & $1.86$ & $-57.9$ & $-57.8$ \\
 & \textit{N},\textit{N}-dimethylpyridin-4-amine & $2.80$ & $-67.3$ & $-67.2$ \\
 & 3-chloropyridine & $1.31$ & $-51.4$ & $-51.2$ \\
 & 2,6-difluoropyridine & $0.14$ & $-41.8$ & $-41.4$ \\
 & 1-methylimidazole & $2.72$ & $-68.0$ & $-67.8$ \\
 & 1-methylpyrazole & $1.84$ & $-56.4$ & $-56.8$ \\
 & oxazole & $1.30$ & $-52.9$ & $-52.7$ \\
\midrule
nitrile & acetonitrile & $0.91$ & $-50.2$ & $-50.3$ \\
 & benzonitrile & $0.80$ & $-49.1$ & $-49.5$ \\
 & 2-chloroacetonitrile & $0.39$ & $-42.3$ & $-42.4$ \\
 & dimethylcyanamide & $1.56$ & $-55.8$ & $-56.1$ \\
 & 2,2-dimethylpropanenitrile & $0.99$ & $-51.6$ & $-52.3$ \\
\midrule
amine & \textit{N},\textit{N}-dimethylmethanamine & $2.13$ & $-57.1$ & $-56.2$ \\
 & \textit{N},\textit{N}-diethylethanamine & $1.98$ & $-57.1$ & $-56.7$ \\
 & quinuclidine & $2.71$ & $-62.1$ & $-61.1$ \\
 & pyrrolidine & $2.59$ & $-62.7$ & $-62.2$ \\
 & 2,2,2-trifluoroethanamine & $0.67$ & $-47.0$ & $-46.0$ \\
\midrule
sulfoxide & methylsulfinylmethane & $2.54$ & $-56.1$ & $-56.9$ \\
 & thiolane 1-oxide & $2.46$ & $-57.5$ & $-58.6$ \\
 & methylsulfinylbenzene & $2.24$ & $-54.3$ & $-54.9$ \\
 & benzenesulfinylbenzene & $2.04$ & $-52.9$ & $-53.2$ \\
 & 1-chloro-4-(4-chlorophenyl)sulfinylbenzene & $1.68$ & $-47.4$ & $-47.4$ \\
\midrule
phosphoryl & dimethylphosphorylmethane & $3.53$ & $-64.4$ & $-65.9$ \\
 & 1-diethylphosphorylethane & $3.66$ & $-66.1$ & $-66.9$ \\
 & [chloro(phenyl)phosphoryl]benzene & $2.20$ & $-55.1$ & $-55.3$ \\
 & dichlorophosphorylbenzene & $1.28$ & $-46.0$ & $-46.5$ \\
 & phosphoryl chloride & $0.56$ & $-36.3$ & $-36.8$ \\
 & triethyl phosphate & $2.68$ & $-55.4$ & $-56.6$ \\
\bottomrule
\end{tabular}
\end{table}

\paragraph{Method.}
The ESP was evaluated on spherical shells from 0.9 to 2.0~\AA{} around the acceptor atom, keeping the points closer to the acceptor than to any other nucleus, and its lowest value was refined by a Nelder--Mead minimization to give $V_\mathrm{min}$.
DFT references were computed at the same geometries with $\omega$B97M-V/def2-TZVPD in GPU4PySCF, for the full density and for its Coulomb fit on \texttt{def2-universal-jfit}.

\paragraph{Results.}
$V_\mathrm{min}$ of \PETrho{} differs from that of the fitted DFT density by 0.49~kcal\,mol$^{-1}$ on average (at most 1.4) and from that of the full DFT density by 0.47~kcal\,mol$^{-1}$ (at most 1.5).
Against p$K_\mathrm{BHX}$, the linear fit over all 44 acceptors gives $r = -0.93$ and an RMSE of 0.34 for \PETrho{}, against $-0.94$ and 0.32 for DFT; within each family, $|r| \geq 0.96$ and the RMSE is at most 0.17 for all three densities.
The residuals of the pooled fit are offsets between families, nearly the same for \PETrho{} and DFT: $+0.5$ for phosphoryl compounds, $+0.4$ for sulfoxides, $-0.4$ for nitriles and $-0.4$ for pyridines and azoles ($-0.3$ for DFT).

\paragraph{Bifunctional bases.}
The p$K_\mathrm{BHX}$ of each site was predicted from its $V_\mathrm{min}$, and the share of the 1:1 complexes at site $i$ was taken as $K_i / \sum_j K_j$.
Each base was computed in a single conformer, the lowest of 20 MMFF conformers relaxed with \texttt{PET-OMol-L}, as for the 44 acceptors.
Table~\ref{tab:hb-sites} compares the predictions with the 12 bases for which we found a measured split, from FTIR in CCl$_4$ against 4-fluorophenol~\cite{graton2003nicotine,graton2001thesis}.
With the pooled fit, the major site is right in 9 of 12.
The three misses are aminopropionitriles, in which the amine is the major site: the pooled line overestimates the p$K_\mathrm{BHX}$ of nitriles by about 0.4, so it puts the major site on the nitrile.

\begin{table}[!htbp]
\centering
\caption{Share of hydrogen-bonded complexes at the major measured site of bifunctional bases: measured~\cite{graton2003nicotine,graton2001thesis} and predicted from the \PETrho{} $V_\mathrm{min}$ with the pooled fit.}
\label{tab:hb-sites}
\begin{tabular}{@{}llcc@{}}
\toprule
Base & Major site & Measured & Predicted \\
\midrule
nicotine & pyridine N & 0.90 & 0.95 \\
nornicotine & pyridine N & 0.80 & 0.93 \\
2-methoxyethylamine & amine N & 0.94 & 0.98 \\
3-methoxypropylamine & amine N & 0.92 & 0.97 \\
morpholine & amine N & 0.83 & 0.89 \\
$N$-methylmorpholine & amine N & 0.77 & 0.80 \\
bis(2-methoxyethyl)amine & amine N & 0.95 & 0.89 \\
$N$-methylaminoacetonitrile & nitrile N & 0.58 & 0.69 \\
$N,N$-dimethylaminoacetonitrile & nitrile N & 0.55 & 0.80 \\
3-aminopropionitrile & amine N & 0.81 & 0.46 \\
$N$-methylaminopropionitrile & amine N & 0.73 & 0.46 \\
$N,N$-dimethylaminopropionitrile & amine N & 0.64 & 0.16 \\
\bottomrule
\end{tabular}
\end{table}

\FloatBarrier
\section{Core-level shifts}
\label{sec:si-xps}

\paragraph{Method.}
The potential at nucleus $I$, excluding its own charge, is
\begin{equation}
V_I = \sum_{J\neq I} \frac{Z_J}{|\mathbf{R}_J - \mathbf{R}_I|} - \int \frac{\rho(\mathbf{r})}{|\mathbf{r} - \mathbf{R}_I|}\,d\mathbf{r},
\end{equation}
which for the predicted density is a sum of two-center Coulomb integrals between the auxiliary functions and a point charge at $\mathbf{R}_I$.
In the equivalent-cores approximation~\cite{jolly1970equivalent}, the final state of the ionization of a 1s electron of atom $I$ is represented by the molecule in which $I$ is replaced by the element with $Z_I+1$, with total charge $+1$.
The transition-potential estimate of the binding-energy shift of site $I$ with respect to a reference is
\begin{equation}
\Delta E_I = \tfrac{1}{2}\left[\Delta V_I^{(0)} + \Delta V_I^{(Z+1)}\right] + c_{Z_I},
\end{equation}
where $V_I^{(0)}$ and $V_I^{(Z+1)}$ are computed on the neutral molecule and on the equivalent-core cation at the same geometry, and $c_{Z_I}$ is one offset per element, equal to the mean difference between computed and measured binding energies of that element.
The initial-state estimate uses $\Delta V_I^{(0)}$ alone.
To check that the auxiliary basis does not limit the shifts, we computed PBE0/def2-TZVPD densities of six molecules and compared the potential at the carbon nuclei from the density matrix with that from its Coulomb-metric fit on \texttt{def2-universal-jfit}: the fit changes the C1s shifts by at most 0.07~eV.

\paragraph{Data and geometries.}
Experimental vertical 1s binding energies of 142 sites (75 C, 22 N, 39 O, 6 F) in 60 closed-shell molecules containing H, C, N, O and F were collected from an earlier compilation~\cite{jolly1984core} as quoted in the CORE65 benchmark~\cite{golze2020core}, from the UT gas-phase XPS database and from two synchrotron studies~\cite{travnikova2012esca,plekan2008thymine}.
The UT database cites the original measurement of each value.
Table~\ref{tab:xps-data} lists the molecules, the elements measured in each and the source of the values.
Where several values exist, high-resolution measurements were preferred.
Unresolved peaks of inequivalent atoms were assigned to one representative site.
Geometries were generated with RDKit (ETKDG and MMFF) and relaxed with \texttt{PET-OMol-L} to a maximum force of 0.01~eV\,\AA$^{-1}$.
DFT densities were computed with PySCF and GPU4PySCF with density fitting at the $\omega$B97M-V/def2-TZVPD level.

\begin{table}[!htbp]
\centering
\footnotesize
\caption{Molecules of the core-level benchmark, with the elements of the measured 1s sites (number of sites in parentheses) and the source of the binding energies. C: CORE65~\cite{golze2020core}; U: UT gas-phase XPS database (\protect\url{https://github.com/UT-XPS/gas-phase-database}), a compilation that cites the original measurement of each value; T: synchrotron study of ethyl trifluoroacetate~\cite{travnikova2012esca}; P: synchrotron study of thymine~\cite{plekan2008thymine}.}
\label{tab:xps-data}
\begin{tabular}{@{}lll@{\hspace{2em}}lll@{}}
\toprule
Molecule & Sites & Source & Molecule & Sites & Source \\
\midrule
ethyl trifluoroacetate & C\,(4), O\,(2) & U, T & 1,1-difluoroethene & C\,(2) & U \\
methane & C & C & hydrogen cyanide & C, N & C \\
ethane & C & C & acetonitrile & C\,(2), N & C \\
ethene & C & C & methylamine & C, N & C, U \\
ethyne & C & C & trimethylamine & N & C \\
carbon monoxide & C, O & C & formamide & C, N, O & U \\
carbon dioxide & C, O & C & acetamide & N & U \\
tetrafluoromethane & C, F & C & \textit{N},\textit{N}-dimethylformamide & C, N, O & U \\
fluoromethane & C, F & C & \textit{N},\textit{N}-dimethylacetamide & N & U \\
trifluoromethane & C, F & C & urea & C, N, O & C \\
difluoromethane & C & U & glycine & C\,(2), N, O\,(2) & C \\
hydrogen fluoride & F & U & nitrous oxide & N\,(2), O & U \\
water & O & C & benzene & C & C \\
ammonia & N & C & pyridine & N & C \\
methanol & C, O & C & pyrrole & C\,(2), N & C, U \\
ethanol & C\,(2), O & U & aniline & N & C \\
dimethyl ether & C, O & C & nitrobenzene & C\,(2), N, O & C \\
formaldehyde & C, O & C & phenylacetylene & C\,(4) & C \\
acetaldehyde & C & U & fluorobenzene & C, F & U \\
acetone & C\,(2), O & C & (trifluoromethyl)benzene & C, F & U \\
formic acid & C, O\,(2) & C & acetophenone & C, O & U \\
acetic acid & C\,(2), O\,(2) & C & benzophenone & C, O & U \\
methyl formate & C\,(2), O\,(2) & C, U & fluorenone & C, O & U \\
methyl acetate & C\,(3), O\,(2) & U & anthrone & C, O & U \\
ethyl acetate & C\,(3), O\,(2) & U & diphenyl carbonate & O & U \\
ethyl fluoroacetate & C\,(4), O\,(2) & U & 1,8-bis(dimethylamino)naphthalene & N & U \\
3,3,3-trifluoropropanoic acid & C\,(3), O\,(2) & U & 1-naphthylamine & N & U \\
trifluoromethyl formate & C\,(2), O\,(2) & U & \textit{N},\textit{N}-diethylaniline & N & U \\
1,1,1-trifluoroethane & C\,(2) & U & quinuclidine & N & U \\
fluoroethane & C\,(2) & U & thymine & C\,(3), N, O & P \\
\bottomrule
\end{tabular}
\end{table}

\paragraph{Results.}
Table~\ref{tab:xps} gives the errors of the initial-state and transition-potential estimates by element.
With these offsets, the initial-state shifts of \PETrho{} are within 0.10~eV of DFT at every site.
The largest errors are the C1s of CO (equivalent core NO$^+$, 2.6~eV), the N1s of acetonitrile (CH$_3$CO$^+$), the O1s of formaldehyde, the F1s of HF (NeH$^+$) and the other F1s sites, whose equivalent cores contain Ne$^+$.
OMol25 contains neon only as neutral atoms in clusters with water and other noble gases, so a bonded Ne$^+$ is outside the training set.

\begin{table}[!htbp]
\centering
\small
\caption{Mean absolute error (eV) of core-level shifts against experiment, with one offset per element, from the DFT density and from \PETrho{}.}
\label{tab:xps}
\begin{tabular*}{\linewidth}{@{\extracolsep{\fill}}lccccc@{}}
\toprule
 & C (75) & N (22) & O (39) & F (6) & all (142) \\
\midrule
Initial state, DFT & 0.46 & 0.59 & 0.40 & 0.41 & 0.46 \\
Initial state, \PETrho{} & 0.46 & 0.60 & 0.40 & 0.41 & 0.46 \\
Transition potential, DFT & 0.16 & 0.22 & 0.14 & 0.26 & 0.17 \\
Transition potential, \PETrho{} & 0.21 & 0.37 & 0.14 & 0.73 & 0.24 \\
\bottomrule
\end{tabular*}
\end{table}

\FloatBarrier
\section{Comparison with cheap methods}
\label{sec:si-baselines}

\paragraph{Baselines.}
We compared \PETrho{} with two approximations to the density or the potential that cost far less than DFT and are used for the same observables.
The density of GFN2-xTB~\cite{bannwarth2019gfn2}, a self-consistent tight-binding method parametrized up to Rn that runs in seconds on these molecules, was built from the occupied orbitals in its molden output.
GFN2-xTB describes only the valence electrons, so its ESP is that of this valence density and of nuclei that carry the valence charge.
The Mulliken charges of GFN2-xTB give the cheapest atomistic model of the potential, a sum of point charges.

\paragraph{Protocol.}
Each baseline was evaluated with the protocol of the corresponding observable in the main text, with two differences.
First, the $\sigma$-hole of the xTB density and charges is evaluated on the $\rho = 0.001$~a.u.\ surface of the xTB density, in which points within 1~\AA{} of a heavy nucleus count as inside, to remove the inner contours of a density without core electrons.
For the surface ESP of the test set, all methods are evaluated on the points of the DFT isosurface of Sec.~\ref{sec:si-esp}.
Second, the potential of point charges has no lone-pair minimum, so its $V_\mathrm{min}$ is the lowest value on the search shell, which lies on its inner or outer boundary.
For core-level shifts, the potential at the nucleus of the xTB density was used in the transition-potential form of Sec.~\ref{sec:si-xps}, with the equivalent-core cation also computed with GFN2-xTB.
For the test set, 40 random structures were drawn from each subset of Fig.~\ref{fig:accuracy}b of the main text (320 in total), of which 4 GFN2-xTB calculations failed.

\paragraph{Surface ESP.}
The median RMSE of the surface ESP against DFT over the 320 structures (316 for xTB) is 1.16~kcal\,mol$^{-1}$\,e$^{-1}$ for \PETrho{}, 5.4 for the xTB density and 8.0 for the xTB charges (Table~\ref{tab:baselines-esp}).
\PETrho{} is 4.1 to 5.2 times closer to DFT than the xTB density in every subset except metal reactivity, where it is 2.7 times closer.

\begin{table}[!htbp]
\centering
\small
\caption{Median RMSE of the surface ESP against DFT (kcal\,mol$^{-1}$\,e$^{-1}$) for 40 random test structures per subset of Fig.~\ref{fig:accuracy}b of the main text.}
\label{tab:baselines-esp}
\begin{tabular}{@{}lccc@{}}
\toprule
Subset & \PETrho{} & xTB density & xTB charges \\
\midrule
small organics & 0.96 & 4.52 & 7.50 \\
organic reactions & 0.97 & 4.01 & 7.99 \\
biomolecules & 1.03 & 4.49 & 7.09 \\
electrolytes & 0.99 & 5.15 & 6.27 \\
scaled separations & 1.27 & 5.40 & 7.91 \\
metal complexes & 1.30 & 6.47 & 7.48 \\
metal reactivity & 3.08 & 8.44 & 10.72 \\
electrolyte redox, reactivity & 2.26 & 10.97 & 14.83 \\
\bottomrule
\end{tabular}
\end{table}

\paragraph{Observables.}
Both xTB baselines follow the Hammett trend of the $\sigma$-hole ($r = 0.93$ and 0.94), but on the cap of iodine their potential is negative, 29 to 37~kcal\,mol$^{-1}$ below that of the DFT density for X = NH$_2$, H and NO$_2$, so neither has a $\sigma$-hole (Table~\ref{tab:baselines-obs}).
For hydrogen-bond basicity, $V_\mathrm{min}$ correlates with p$K_\mathrm{BHX}$ at $r = -0.94$ for DFT and $-0.93$ for \PETrho{}, against $-0.77$ for the xTB density and $-0.46$ for the xTB charges.
The $V_\mathrm{min}$ of the xTB density is 42~kcal\,mol$^{-1}$ too negative on average, against 0.47~kcal\,mol$^{-1}$ for \PETrho{}.
On the major site of the bifunctional bases, however, the xTB density does as well as \PETrho{}: it is right in 10 of 12, against 9 for \PETrho{}.
The share of complexes depends only on the difference between the two minima of one molecule, in which a common offset of the xTB density likely cancels.
For core-level shifts, the xTB density in the transition-potential form gives 1.93~eV, against 0.24~eV for \PETrho{} and 0.17~eV for DFT.

\begin{table}[!htbp]
\centering
\small
\caption{Observables of Fig.~\ref{fig:observables}b--d of the main text from DFT, \PETrho{} and the cheap baselines: correlation of the $\sigma$-hole maximum with $\sigma_p$ (20 iodobenzenes) and its difference from the DFT density for X = NH$_2$, H and NO$_2$ (kcal\,mol$^{-1}$); correlation of $V_\mathrm{min}$ with p$K_\mathrm{BHX}$ (44 acceptors), its mean absolute difference from DFT (kcal\,mol$^{-1}$) and the number of bifunctional bases (of 12) with the right major site; mean absolute error of core-level shifts against experiment (eV, 142 sites).}
\label{tab:baselines-obs}
\begin{tabular*}{\linewidth}{@{\extracolsep{\fill}}lcccccc@{}}
\toprule
 & \multicolumn{2}{c}{$\sigma$-hole} & \multicolumn{3}{c}{Hydrogen-bond basicity} & Core levels \\
\cmidrule(lr){2-3}\cmidrule(lr){4-6}\cmidrule(l){7-7}
 & $r$ & $\Delta V_{S,\max}$ & $r$ & MAE & Major site & MAE \\
\midrule
DFT & 0.96 & -- & $-0.94$ & -- & -- & 0.17 \\
\PETrho{} & 0.96 & $+0.4$ to $+1.0$ & $-0.93$ & 0.47 & 9 & 0.24 \\
xTB density & 0.93 & $-37$ to $-35$ & $-0.77$ & 42 & 10 & 1.93 \\
xTB charges & 0.94 & $-30$ to $-29$ & $-0.46$ & 30 & 4 & -- \\
\bottomrule
\end{tabular*}
\end{table}

\FloatBarrier
\section{Charge and spin conditioning}
\label{sec:si-charge-spin}

\paragraph{Same-geometry pairs.}
OMol25 contains 104 pairs of structures with identical geometry and different charge (4) or spin multiplicity (100) in which one structure is in the validation set; for all but one, the partner is in the training set, and the remaining partner is in the test set.
For each pair, we compare the predicted difference density $\Delta\rho = \rho_B - \rho_A$, with each density normalized to its electron count, with that of the fitted reference and report the relative error $\int|\Delta\rho_\text{pred} - \Delta\rho_\text{ref}| / \int|\Delta\rho_\text{ref}|$ on a PySCF integration grid (level 3), which is 1 for a model that ignores the embedding (Table~\ref{tab:pairs}).
A model that predicts one state and only rescales it to the electron count of the other gives 1 for spin pairs, where the counts are equal, and 1.64 for charge pairs.
The relative error counts misplaced density twice, once where it is missing and once where it is in excess, so we also report the fraction of the added density that the model places where the reference places it,
\begin{equation}
f_+ = \frac{\int \max\!\left[\min(\Delta\rho_\text{pred}, \Delta\rho_\text{ref}), 0\right] d\mathbf{r}}{\int \max(\Delta\rho_\text{ref}, 0)\, d\mathbf{r}},
\end{equation}
which is 1 when the predicted gain covers the reference gain everywhere, and the same fraction $f_-$ for the removed density, with $-\Delta\rho$ in place of $\Delta\rho$.
Here $\Delta\rho$ is the density of the state with lower charge or higher multiplicity minus that of its partner.
The medians of $f_+$ ($f_-$) are 0.81 (0.83) for spin pairs and 0.71 (0.37) for charge pairs.

\begin{table}[!htbp]
\centering
\small
\caption{Relative error of the difference density between two states at the same geometry: median and interquartile range.}
\label{tab:pairs}
\begin{tabular}{@{}lcc@{}}
\toprule
 & pairs & \PETrho{} \\
\midrule
spin & 100 & 0.33 (0.21--0.53) \\
charge & 4 & 0.55 (0.48--0.63) \\
\bottomrule
\end{tabular}
\end{table}

\paragraph{Separated molecules.}
For the anion of tetracyanoethylene (TCNE) and benzoquinone 10~\AA{} apart, the difference between the densities of the anion and of the neutral pair integrates to 0.62~e on TCNE over Becke cells with \PETrho{}, against a whole electron in DFT.
For the cation of three 1,2-dibromoethane molecules, the difference between the densities of the neutral cluster and of the cation integrates to 0.98~e on one molecule and 0.01~e on each of the other two in DFT, and to 0.25, 0.39 and 0.36~e with \PETrho{}.

\begin{figure}[!htbp]
    \centering
    \includegraphics{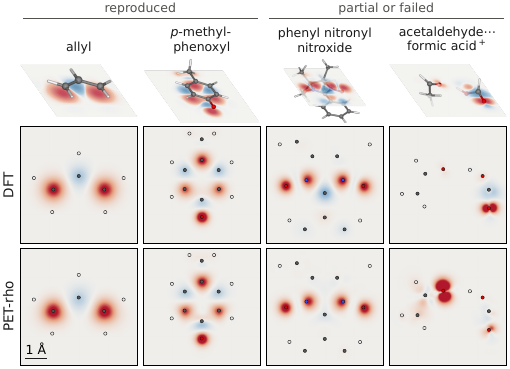}
    \caption{Spin density of the allyl, $p$-methylphenoxyl and phenyl nitronyl nitroxide radicals and of an acetaldehyde--formic acid radical cation, from DFT (top) and \PETrho{} (bottom) in one plane per system, shown in the render above each column.
    Color scales span $\pm$ the 99.5th percentile of the absolute DFT map (red positive, blue negative): from left to right, 0.064, 0.041, 0.057 and 0.037~e\,bohr$^{-3}$.
    Dots mark atoms within 0.9~\AA{} of the plane.}
    \label{fig:si-spin-maps}
\end{figure}

\paragraph{Spin populations.}
Over the 5,893 open-shell test structures with up to 100 atoms, the spin density of \PETrho{} integrated over Becke atoms is within 0.013~e of DFT on average, the sign is correct for 97\% of the atoms that carry more than 0.02~e, and the moments of transition-metal atoms are within 4\% (median).
The negative populations produced by spin polarization are recovered at about a third of their magnitude, with the correct sign for two atoms in three, and the error is as large on training structures.
They are correct in allyl, whereas for the central carbon of phenyl nitronyl nitroxide the prediction is $-0.03$~e, against $-0.16$~e in DFT (Fig.~\ref{fig:si-spin-maps}).
On 244 open-shell test structures with 592 atoms below $-0.02$~e, Becke populations of the DFT spin density matrix, of the fitted spin density and of the prediction were computed on the same grid (level 3).
The fit reproduces the negative DFT populations exactly (ratio 1.000, all signs correct), so the prediction has the same error against DFT as against the fit.
The fraction recovered by \PETrho{} falls with the size of the population, from a median of 0.40 between $-0.10$ and $-0.05$~e to 0.06 below $-0.10$~e, where half of the signs are wrong.

\paragraph{Training data.}
We searched the four million OMol25 structures for the bond graphs of the systems of Fig.~\ref{fig:spin} of the main text and of Fig.~\ref{fig:si-spin-maps} at any geometry, as whole structures and as connected components of larger structures (Table~\ref{tab:fig3-train}).
Two atoms are bonded when their distance is below 1.2 times the sum of their covalent radii.
The Fe(H$_2$O)$_6^{2+}$ singlet, the tetrathiafulvalene (TTF) cation, the TCNE--benzoquinone pair and the three radicals do not occur as whole structures in any split.
The dibromoethane cluster and the acetaldehyde--formic acid cation are themselves test structures, and the training set contains other geometries of both in the same charge and multiplicity.

\begin{table}[!htbp]
\centering
\small
\caption{OMol25 structures with the bond graph of the systems of Fig.~\ref{fig:spin} of the main text and of Fig.~\ref{fig:si-spin-maps}: whole structures in the same charge $q$ and multiplicity $2S+1$, by split, and connected components of larger training structures, in any state.}
\label{tab:fig3-train}
\begin{tabular}{@{}lcrrrr@{}}
\toprule
System & $q$, $2S+1$ & Train & Val. & Test & Component \\
\midrule
Fe(H$_2$O)$_6^{2+}$ & 2, 1 & 0 & 0 & 0 & 33 \\
Fe(H$_2$O)$_6^{2+}$ & 2, 5 & 29 & 0 & 1 & \\
TTF & 0, 1 & 6 & 0 & 0 & 0 \\
TTF$^+$ & 1, 2 & 0 & 0 & 0 & \\
TCNE--benzoquinone & 0, 1 & 0 & 0 & 0 & 5, 6596$^a$ \\
TCNE--benzoquinone$^-$ & $-1$, 2 & 0 & 0 & 0 & \\
(C$_2$H$_4$Br$_2$)$_3$ & 0, 1 & 52 & 0 & 0 & 13584$^b$ \\
(C$_2$H$_4$Br$_2$)$_3^+$ & 1, 2 & 16 & 0 & 1 & \\
allyl & 0, 2 & 0 & 0 & 0 & 908 \\
$p$-methylphenoxyl & 0, 2 & 0 & 0 & 0 & 303 \\
phenyl nitronyl nitroxide & 0, 2 & 0 & 0 & 0 & 0 \\
acetaldehyde--formic acid$^+$ & 1, 2 & 12 & 1 & 1 & 0 \\
\bottomrule
\end{tabular}

\raggedright\footnotesize
$^a$TCNE and benzoquinone. $^b$1,2-Dibromoethane.
\end{table}

\FloatBarrier
\FloatBarrier
\section{Error estimates}
\label{sec:si-uq}

\paragraph{Ensemble.}
The readout of \PETrho{} is linear: in each block of the coefficients of one angular momentum $\lambda$ of one element $Z$, the coefficients of an atom are $W\mathbf{f}$, where $\mathbf{f}$ holds the $d = 1536$ last-layer features of the atom (1024 from the atom head and 512 from the summed pair head) and $W$ is the $n_\mathrm{out} \times d$ weight matrix of the block.
In the last-layer prediction rigidity (LLPR)~\cite{bigi2024llpr,chong2025}, $W$ is viewed as the solution of a linear regression on the training set, with precision matrix
\begin{equation}
\Sigma_{Z\lambda} = F^\top F + \varsigma^2 I, \qquad \varsigma^2 = \frac{\operatorname{tr}(F^\top F)}{d\,N_Z},
\end{equation}
where the rows of $F$ are the features of the $N_Z$ training atoms of element $Z$ and the regularizer $\varsigma^2$ is the mean squared feature.
Member $k$ of an ensemble of 32 perturbs the weights of each block by
\begin{equation}
\Delta W_k = L\,\varepsilon_k\,\Sigma_{Z\lambda}^{-1/2},
\end{equation}
where $\varepsilon_k$ is an $n_\mathrm{out} \times d$ matrix of independent standard normal numbers and $L L^\top$ is the pseudo-inverse of the Coulomb metric of the coefficients of the block, so that the deviations are drawn in the Coulomb metric of the auxiliary basis, as in the loss.
We sample an ensemble instead of propagating the covariance because the error we predict, the integral of $|\rho_\text{pred} - \rho_\text{ref}|$, is nonlinear in the coefficients.
The charge normalization removes from the coefficient deviation $\mathbf{d}_k$ of each member its change of electron count, through the projector $P\mathbf{d} = \mathbf{d} - \mathbf{u}\,(\mathbf{S}\cdot\mathbf{d})/(\mathbf{S}\cdot\mathbf{u})$, where $\mathbf{S}$ holds the integrals of the auxiliary functions, $\mathbf{u} = J^{-1}\mathbf{S}$ and $J$ is the Coulomb metric of the auxiliary basis.
The mean deviation over the members is subtracted from each, so that the mean of the ensemble is the prediction itself.

\paragraph{Calibration.}
In the original LLPR, all deviations are multiplied by one factor $\alpha$, chosen so that the spread of the ensemble matches the residuals on a validation set~\cite{bigi2024llpr}.
The scale of $\Sigma_{Z\lambda}^{-1}$, however, depends on the regularizer $\varsigma^2$ and on the training errors of the block, and both differ between elements and angular momenta, so that with a single factor the error of some chemistries is overestimated and that of others underestimated (paragraph Scalings below).
We therefore scale the deviation of each group $g = (Z, \lambda)$ of auxiliary functions by its own factor,
\begin{equation}
    P\mathbf{d}_k = \sum_g P\mathbf{d}_{kg} \;\longrightarrow\; \sum_g \alpha_g\, P\mathbf{d}_{kg},
\end{equation}
where $Z$ is the atomic number, also for elements with an effective core potential.
The factors are fitted on the Coulomb self-energy of the error, the quantity minimized in training, because it splits over the groups: the perturbations of different blocks are drawn independently, so the cross terms vanish on average and the expected Coulomb error of a structure is
\begin{equation}
    E^\mathrm{C}_\mathrm{pred} = \sum_g \alpha_g^2 E^\mathrm{C}_g, \qquad E^\mathrm{C}_g = \Big\langle \tfrac{1}{2}\, (P\mathbf{d}_{kg})^\top J\, (P\mathbf{d}_{kg}) \Big\rangle_k .
\end{equation}
Since $E^\mathrm{C}_\mathrm{pred}$ is linear in the $\alpha_g^2$, a single evaluation of the unscaled ensemble on the validation set, which gives the $E^\mathrm{C}_g$ of every structure, suffices to fit all factors.
With $\log \alpha_{Z\lambda}^2 = t_Z + \delta_{Z\lambda}$, the factors are fitted by minimizing
\begin{equation}
    \sum_s \left[\log E^\mathrm{C}_{\mathrm{obs},s} - \log E^\mathrm{C}_{\mathrm{pred},s}\right]^2 + \mu_\delta \sum_{Z\lambda} \delta_{Z\lambda}^2 + \mu_t \sum_Z (t_Z - \bar t)^2 ,
\end{equation}
where $E^\mathrm{C}_{\mathrm{obs},s} = \frac{1}{2}\mathbf{e}_s^\top J \mathbf{e}_s$ is the Coulomb error of structure $s$, $\mathbf{e}_s$ the difference between its reference and predicted coefficients, $\bar t$ the mean of the $t_Z$ over the elements, and $\mu_\delta = \mu_t = 0.1$.
The penalties shrink each $\alpha_{Z\lambda}$ towards a common factor for its element and the element factors towards each other, so that blocks seen in few validation structures stay close to their element.
The least-squares fit on the logarithm calibrates the median of the observed-to-predicted ratio and keeps the few structures with errors far above the prediction from dominating the scale, as they would in a fit of the squared errors themselves.
A factor that multiplies a whole block of angular momentum $\lambda$ preserves the equivariance of the prediction.

\paragraph{Predicted error.}
The predicted density error of a structure is the mean over the members, scaled by $\alpha_{Z\lambda}$, of the error of the member density against the prediction, computed on the same grid as the observed one.
Scaling a block by $\alpha_g$ scales its deviation by $\alpha_g$, so the factors fitted on the Coulomb error apply to any error measure computed from the member densities.
The observed and the predicted error are divided by the same number of atoms, so their ratio, and every coverage derived from it, is the same per atom and per electron.
On the test set, predicted and observed errors rank-correlate at 0.91.

\paragraph{Bands.}
For a given structure, the NMAEs of the members agree to about 10\%, while the observed NMAE scatters around the prediction by about a factor of two.
The NMAE integrates many nearly independent directions of the Gaussian ensemble, so its spread over the members is small, whereas the actual errors are concentrated in a few chemically defined modes, such as a misplaced charge or spin.
The ensemble therefore predicts the magnitude of the error but not its distribution, and the bands of Fig.~\ref{fig:trust} of the main text are not derived from a Gaussian as in the original LLPR~\cite{bigi2024llpr}.
They are the 0.135, 2.5, 16, 84, 97.5 and 99.865\% quantiles of $\log(\mathrm{NMAE}^\mathrm{obs}/\mathrm{NMAE}^\mathrm{pred})$ on the validation set, shifted to zero median and drawn around the diagonal, labelled 1$\sigma$, 2$\sigma$ and 3$\sigma$ after the Gaussian coverage they correspond to; for the density they contain 68, 95 and 99.7\% of the test structures.

\begin{figure*}[!htbp]
    \centering
    \includegraphics{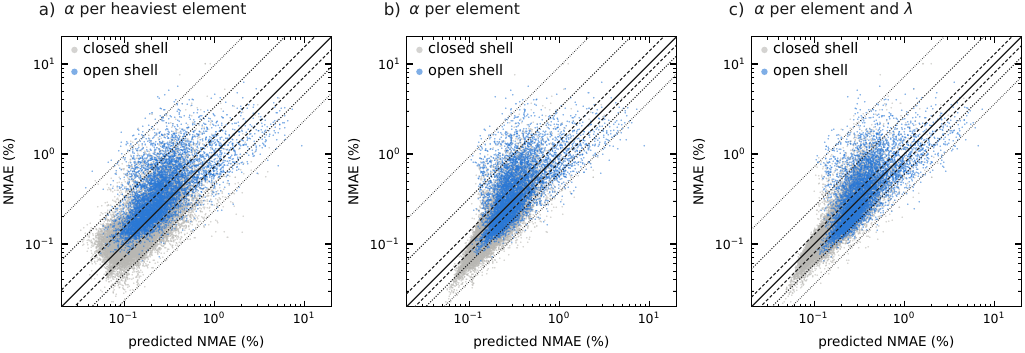}
    \caption{%
    Observed against predicted NMAE of \PETrho{} on the test set for three scalings of the ensemble, closed (grey) and open (blue) shells; lines and bands as in Fig.~\ref{fig:trust} of the main text.
    (a)~One factor per heaviest element, (b)~one factor per element and (c)~one factor per element and $\lambda$, as in Fig.~\ref{fig:trust} of the main text.
    The bands are fitted on the validation set separately for each scaling.}
    \label{fig:si-trust-alpha}
\end{figure*}

\paragraph{Scalings.}
Of the three scalings of the ensemble in Fig.~\ref{fig:si-trust-alpha}, all fitted in the same way, the one with one factor per heaviest element of the structure underestimates the error of open shells (median ratio of observed to predicted NMAE 1.34, against 1.00 for closed shells).
The rank correlation between predicted and observed errors is 0.77, against 0.91 with one factor per element and angular momentum, which also gives narrower bands.

\begin{figure}[!htbp]
    \centering
    \includegraphics{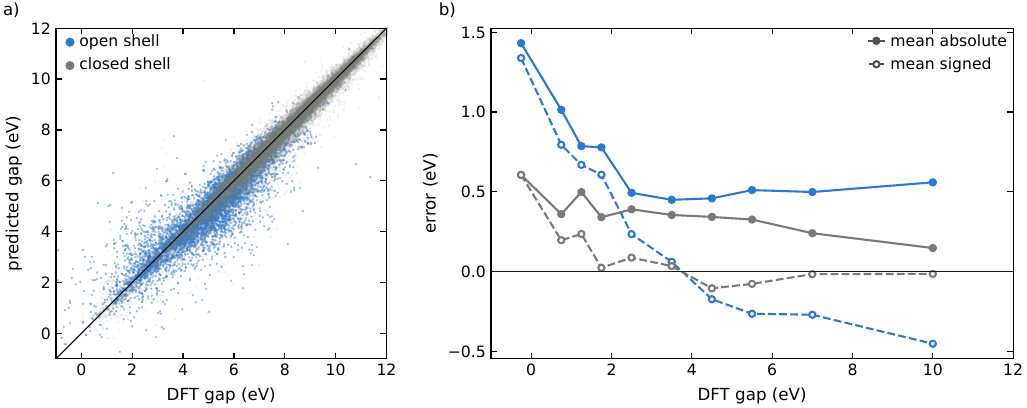}
    \caption{HOMO--LUMO gap predicted by the gap head on the frozen backbone of \PETrho{} for all 39,865 test structures, noble gases included, closed (grey) and open (blue) shells: (a) against the DFT gap; (b) mean absolute (filled, solid) and mean signed (open, dashed) error in bins of the DFT gap.}
    \label{fig:si-gap}
\end{figure}

\paragraph{Gap head.}
The head for the HOMO--LUMO gap was trained on the frozen backbone of \PETrho{}.
It predicts two scalars per atom, whose softmax-weighted maximum and minimum over the structure give the highest occupied and lowest unoccupied levels~\cite{malosso2026transferable}; their difference is the gap.
The reference is the gap of OMol25 between the lowest unoccupied and the highest occupied orbital over both spin channels.
Gaps below 2~eV are 1.6\% of the training set, so the training set of the head contains all training structures with gaps below 4~eV (302,250), 150,000 random others and two more copies of those below 2~eV (576,472 samples in total); its validation set contains 20,000 random validation structures and all those below 2~eV (20,356 in total, of which 11 are noble-gas structures).
Only the two per-atom heads were trained (2.6 million parameters), for 40 epochs with a Huber loss ($\delta = 0.47$~eV), keeping the epoch with the lowest validation loss.
On the full test set (Fig.~\ref{fig:si-gap}), the mean absolute error is 0.25~eV, 0.20~eV for closed-shell and 0.51~eV for open-shell structures, against 0.13~eV for the original model trained on a single system~\cite{malosso2026transferable}.
The error grows for small gaps, which the head overestimates (by 0.68~eV on average below 1~eV), and the open-shell gaps are compressed towards the mean.

\begin{figure}[!htbp]
    \centering
    \includegraphics{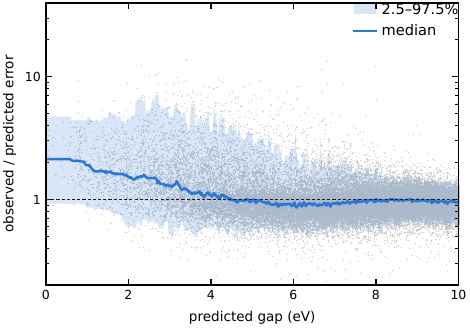}
    \caption{Ratio of observed to LLPR-predicted density NMAE of the 39,845 validation structures against the gap predicted by the gap head (grey), with the median (line) and the 2.5--97.5\% quantiles (band) of the ratios of the 200 validation structures nearest in predicted gap.}
    \label{fig:si-gap-calib}
\end{figure}

\paragraph{Gap calibration.}
The ratio $\mathrm{NMAE}^\mathrm{obs}/\mathrm{NMAE}^\mathrm{pred}$ of the validation structures depends on the gap (Fig.~\ref{fig:si-gap-calib}): its median is about 2 below 1~eV, falls steadily up to about 4.5~eV and stays close to 1 above.
Within a class of the predicted gap, the LLPR still orders the structures by their error, with slopes of $\log \mathrm{NMAE}^\mathrm{obs}$ against $\log \mathrm{NMAE}^\mathrm{pred}$ of about 0.8 below 4~eV and 1.0 above (Table~\ref{tab:gap-classes}).
The estimate is therefore mainly off by a factor that depends on the gap, and we correct it by that factor.

For a structure with predicted gap $g$, the corrected estimate is the LLPR estimate times the median of $\mathrm{NMAE}^\mathrm{obs}/\mathrm{NMAE}^\mathrm{pred}$ over the $k = 200$ validation structures whose predicted gap is closest to $g$, out of all 39,845.
The 2.5 and 97.5\% quantiles of the same ratios give a 95\% band.
This is the calibration of the bands of Fig.~\ref{fig:trust} of the main text, done locally in the gap, and a nearest-neighbor form of localized conformal prediction~\cite{guan2023localized,angelopoulos2023conformal}.
With a fixed number of neighbors, the window is narrow where gaps are common and wide where they are rare; negative predicted gaps are set to zero.
The factor is 2.13 at 0.5~eV, 1.91 at 1~eV, 1.54 at 2~eV, 1.09 at 4~eV and 0.91--0.95 at 6--8~eV.

For the 493 test structures with a DFT gap below 2~eV, the median of $\mathrm{NMAE}^\mathrm{obs}/\mathrm{NMAE}^\mathrm{pred}$ falls from 1.82 to 1.06 with the correction (from 2.44 to 1.26 for the 95 below 1~eV), and the 95\% band contains 89\% of them, against 70\% for the global band of the validation set.
With the 95\% band of each structure, the coverage of the classes of predicted gap below 4~eV rises from 52--84\% to 85--94\% (Table~\ref{tab:gap-classes}).
Below 1~eV the band covers 85\% of the 153 test structures, less than the nominal 95\%: the gaps of the nearest neighbors of such a structure are mostly larger than its own, so the factor and the band are biased towards those of larger gaps.
The validation set also served to fit the factors $\alpha_{Z\lambda}$ and to select the epoch of the gap head, so its ratios are not strictly independent of the fit; the coverages above are therefore given on the test set, which was used for neither.

\begin{table}[!htbp]
\centering
\caption{Classes of the predicted gap: number of validation and test structures, coverage (\%) of the test structures by the global 95\% band of the validation set and by the local 95\% band of the gap calibration, and rank correlation and slope of $\log \mathrm{NMAE}^\mathrm{obs}$ against $\log \mathrm{NMAE}^\mathrm{pred}$ on the test set.}
\label{tab:gap-classes}
\begin{tabular}{@{}lcccccc@{}}
\toprule
Predicted gap (eV) & val & test & global band & local band & rank corr. & slope \\
\midrule
$< 1$ & 157 & 153 & 52 & 85 & 0.74 & 0.79 \\
1--2 & 447 & 434 & 79 & 92 & 0.67 & 0.78 \\
2--3 & 959 & 976 & 80 & 94 & 0.62 & 0.80 \\
3--4 & 1854 & 1785 & 84 & 94 & 0.67 & 0.81 \\
$> 4$ & 36428 & 36497 & 96 & 94 & 0.92 & 1.01 \\
\bottomrule
\end{tabular}
\end{table}

\begin{figure*}[!htbp]
    \centering
    \includegraphics{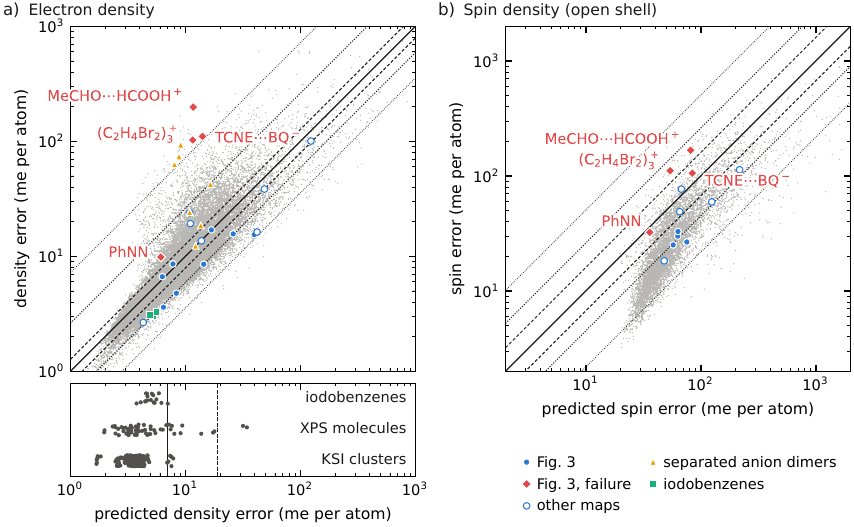}
    \caption{Observed against LLPR-predicted error of \PETrho{} for the systems of Fig.~\ref{fig:spin} of the main text and Fig.~\ref{fig:si-spin-maps}, other spin maps, separated anion dimers and iodobenzenes, on the test set (grey) with the validation bands of Fig.~\ref{fig:trust}: (a)~density error per atom; (b)~spin density error per atom of open-shell structures.
    Failures of Figs.~\ref{fig:spin} and \ref{fig:si-spin-maps} in red.
    Strip below (a): predicted error of the systems whose observed density error was not computed, with the median (solid) and 90th percentile (dashed) of the test set.}
    \label{fig:si-llpr-obs}
\end{figure*}

\paragraph{Systems of Figs.~\ref{fig:observables} and \ref{fig:spin} and of Fig.~\ref{fig:si-spin-maps}.}
Table~\ref{tab:showcase-flag} compares the predicted density error with the error of each observable for all the systems of Fig.~\ref{fig:observables} of the main text with a DFT reference per structure.
The KSI clusters, the hydrogen-bond acceptors and the iodobenzenes lie in the bulk of the test set (Fig.~\ref{fig:si-llpr-obs}), where the predicted error does not rank the structures by the error of the observable.
These errors likely depend on local details of the density that the error per atom does not resolve.
Only the core-level structures reach beyond the 90th percentile: 33 of the 142 equivalent-core cations, of which the six with Ne in place of F lie above the 99th, and the molecules CO, N$_2$O and HCN, whose sites all have a flagged cation.
The 33 flagged sites have a mean error against experiment of 0.49~eV, against 0.16~eV for the other 109, and include 8 of the 10 largest core-level errors.

\begin{table}[htbp]
\centering
\caption{Predicted density error against the error of the observables of Fig.~\ref{fig:observables} of the main text: number of structures, number above the 90th percentile of the test set, highest percentile, Spearman correlation between predicted density error and absolute error of the observable, and mean absolute error of the observable for the quarter of structures with the largest predicted error and for the rest. Observables: field on the C=O bond of the 36 KSI active-site clusters and of their 144 sub-clusters without Tyr16, without Asp103, or with only one of the two (MV\,cm$^{-1}$), $V_\mathrm{min}$ of the acceptors and $\sigma$-hole maximum of the 20 iodobenzenes (kcal\,mol$^{-1}$), and core-level shift, with the larger predicted error of the molecule and of the equivalent-core cation (eV).}
\label{tab:showcase-flag}
\begin{tabular}{lrrrrrr}
\toprule
 & $n$ & $>$P90 & max pct. & Spearman & top 25\% & rest \\
\midrule
KSI clusters, vs DFT & 180 & 0 & 52 & 0.08 & 1.2 & 1.4 \\
acceptors, vs DFT & 44 & 0 & 50 & 0.12 & 0.66 & 0.40 \\
iodobenzenes, vs DFT & 20 & 0 & 20 & 0.10 & 0.88 & 0.79 \\
core-level sites, vs DFT & 142 & 33 & 100 & 0.30 & 0.35 & 0.08 \\
core-level sites, vs experiment & 142 & 33 & 100 & 0.33 & 0.48 & 0.16 \\
\bottomrule
\end{tabular}
\end{table}
The structures of Figs.~\ref{fig:spin} and \ref{fig:si-spin-maps} in which charge or spin is misplaced between molecules are predicted below the 90th percentile.
Their observed density error is beyond the 2$\sigma$ band for the TCNE--benzoquinone anion, the dibromoethane cation and the acetaldehyde--formic acid cation, whose fragments each resemble training structures: the ratio of observed to predicted error is 7.9, 8.9 and 17.0, against 1.6 for phenyl nitronyl nitroxide (PhNN) and 0.56--0.57 for the neutral partners of the two charged pairs.

 \end{document}